\documentclass{aa}  %[dvipdfmx][linenumbers]
\usepackage{graphicx, longtable, comment}
\usepackage{caption, subcaption}
\usepackage{txfonts, lineno}
\newcommand{\tabref}[1]{Tab.~\ref{#1}}
\newcommand{\Tabref}[1]{Tab.~\ref{#1}}

\newcommand{\figref}[1]{Fig.~\ref{#1}}
\newcommand{\Figref}[1]{Fig.~\ref{#1}}
\newcommand{\Figsref}[1]{Figs.~\ref{#1}}
\nolinenumbers
\begin{document}

   \title{Chromospheric Mg I Emission Lines of Pre-Main-Sequence Stars
   \thanks{Just to show the usage of the elements in the author field}
   }
%   \subtitle{I. Overviewing the $\kappa$-mechanism}

   \author{Mai Yamashita\inst{1}, Yoichi Itoh\inst{2}, 
          \and Yuhei Takagi\inst{3}\fnmsep
          }

   \institute{Institute of Space and Astronautical Science, Japan Aerospace Exploration Agency, 3-1-1 Yoshinodai, Chuo-ku, Sagamihara, Kanagawa 252-5210, Japan\\
              \email{yamashita.mai@jaxa.jp}
         \and
             Nishi-Harima Astronomical Observatory, Center for Astronomy, University of Hyogo, 407-2 Nishigaichi, Sayo, Sayo, Hyogo 679-5313, Japan %\\
         \and
             Subaru Telescope, National Astronomical Observatory of Japan, 650 North A'ohoku Place, Hilo, HI 96720, U.S.A.\\
             }

   \date{Received August 28, 2024; accepted September 30, 2024}

% \abstract{}{}{}{}{}
% 5 {} token are mandatory
 
  \abstract
  % context heading (optional)
  % {} leave it empty if necessary  
   { To reveal details of the internal structure, the relationship between chromospheric activity and the Rossby number has been extensively examined for main-sequence stars. For active pre-main sequence (PMS) stars, it is suggested that the level of activity be assessed using optically thin emission lines, such as Mg I.
   }
  % aims heading (mandatory)
   {We aim to detect Mg I chromospheric emission lines from PMS stars and determine whether the chromosphere is activated by the dynamo process or by mass accretion from protoplanetary disks.}
  % methods heading (mandatory)
   { We analyzed high-resolution optical spectra of $64\ $PMS stars obtained with Very Large Telescope (VLT)/X-shooter and UVES and examined the infrared Ca II ($8542\ \mathrm{\AA}$) and Mg I ($8807\ \mathrm{\AA}$) emission lines. To detect the weak chromospheric emission lines, we determined the atmospheric parameters ($T_{\rm eff}$ and $\log\ g$) and the degree of veiling of the PMS stars by comparing the observed spectra with photospheric model spectra.}
  % results heading (mandatory)
   { After subtracting the photospheric model spectrum from the PMS spectrum, we detected Ca II and Mg I as emission lines. The strengths of the Mg I emission lines in PMS stars with no veiling are comparable to those in zero-age main-sequence (ZAMS) stars if both types of stars have similar Rossby numbers. The Mg I emission lines in these PMS stars are thought to be formed by a dynamo process similar to that in ZAMS stars. In contrast, the Mg I emission lines in PMS stars with veiling are stronger than those in ZAMS stars. These objects are believed to have protoplanetary disks, where mass accretion generates shocks near the photosphere, heating the chromosphere.}
  % conclusions heading (optional), leave it empty if necessary
   { The chromosphere of PMS stars is activated not only by the dynamo process but also by mass accretion.}


   \maketitle
%
%________________________________________________________________

\section{Introduction}
The atmosphere above the photosphere is called the chromosphere. The chromosphere is optically thin, with the optical depth gradually decreasing from $\tau_{5000\ \mathrm{\AA}} \approx 10^{-4}$ to $\approx 10^{-7}$ as altitude increases \citep{f04}. In an active chromospheric region, atoms emit permitted lines such as H$\rm \alpha$ and Ca I\hspace{-.1em}I. Chromospheric activity is thought to be driven by the magnetic field generated by the dynamo process. 

For pre-main sequence (PMS) stars, several researchers have suggested that chromospheric emission lines are also influenced by mass accretion. \citet{hp92} conducted optical spectroscopy of PMS stars and found that narrow emission lines, such as Ca II and Mg I (e.g., $8807\ \mathrm{\AA}$), are generated in the stellar chromosphere, while broad emission lines result from mass accretion.  
\citet{mo05} investigated the chromospheric activity of classical T Tauri stars (CTTSs), very low-mass young stars ($0.075 \leq M_* <  0.15 \  \mathrm{M_{\odot}}$), and young brown dwarfs ($M_* \leq 0.075 \  \mathrm{M_{\odot}}$). The surface flux of the Ca II emission line at $8662 \  \mathrm{\AA}$, $F^{\prime}_{8662}$, exhibit correlation with the mass accretion rate, $\dot{M}$, for approximately 4 orders of magnitude. This led them to conclude that the Ca II emission line is an excellent quantitative measure of the accretion rate. \citet{bb96} found correlations with veiling and near-infrared excesses, attributing the Ca II and He I emission lines to a "hot chromosphere" heated by accretion shocks.

\citet{y20} investigated the relationship between the Rossby number, $N_{\rm R}$(=  defined as the rotational period $P$ divided by the convective turnover time $\tau_{\rm c}$), and $R^{\prime}_{\rm 8498}$, $R^{\prime}_{\rm 8542}$, and $R^{\prime}_{\rm 8662}$, which represent the ratio of the surface flux of the Ca II infrared triplet (IRT) emission lines ($8498, 8542, 8662\ \mathrm{\AA}$) to the stellar bolometric luminosity, for $60$ PMS stars. Only three PMS stars exhibited broad and strong emissions, indicative of significant mass accretion. Most PMS stars showed narrow and weak emissions, suggesting that their emission lines are formed in the chromosphere. All their Ca II IRT emission lines had $R^{\prime}_{\rm IRT} \sim 10^{-4.2}$, comparable to the maximum $R^{\prime}_{\rm IRT}$ observed in ZAMS stars. The PMS stars exhibited $N_{\rm R} < 10^{-0.8}$ and a constant $R^{\prime}_{\rm IRT}$ with respect to $N_{\rm R}$, indicating that their Ca II IRT emission lines are saturated.

\citet{y22} demonstrated that Mg I emission lines are a good indicator of activity in fast-rotating and active ZAMS stars. They examined the infrared Mg I emission lines at $8807\ \mathrm{\AA}$ for $47\ $ ZAMS stars in IC 2391 and IC 2602 using archive data from the University College London Echelle Spectrograph on the Anglo-Australian Telescope. They found that ZAMS stars with smaller Rossby numbers exhibit stronger Mg I emission lines, even among stars in the Ca II saturated region.

In this study, we measure the strength of Mg I emission lines in $64\ $PMS stars and investigate whether their chromospheres are activated by the dynamo process or mass accretion. We compare the strength of the chromospheric emission lines in PMS stars with that in ZAMS stars from young open clusters. In the following section, we describe the high-resolution spectroscopic data and the reduction procedures. Sect. \ref{result} presents the results, and Sect. \ref{discussion} discusses the origin of the Mg\ I emission lines and their emitting regions on the stellar surface.

%\newpage
\section{Data sets and data reduction}
\subsection{Data sets} \label{sec:hr}
Our targets include $64\ $G, K, and M-type PMS stars from four star-forming regions and seven moving groups: the Taurus-Auriga molecular cloud, ${\rm\rho}$ Ophiuchi molecular cloud, Lupus star-forming region, Chamaeleon I star-forming region, the Orionis OB 1c association, Upper Scorpius association, AB Doradus moving group, ${\rm\beta}$ Picoris moving group, "Cha-Near" region, ${\rm\eta}$ Chamaeleontis cluster, and TW Hydrae association. Hereafter, objects from these star-forming regions and moving groups are referred to as PMS stars. The metallicity of PMS stars in these regions and groups has been determined to be similar to that of the Sun. We did not include binaries or triplets listed in \citet{ghez93}, \citet{k09}, \citet{k12}, \citet{ne95}, \citet{l93}, \citet{wa10}, and \citet{zs04}. All targets investigated in this study are listed in \tabref{tab:objects1}. % and \ref{tab:objects2}. 

%{\tiny \input{./tables/tabA}}
%\onecolumn
%\input{longtables/test_tabA}
%\twocolumn

We used archival data of $59\ $PMS stars obtained with X-shooter ($R \sim 8,000$) mounted on the Very Large Telescope (VLT). The program IDs, observation dates, and integration times are listed in \tabref{tab:arc}. The wavelength coverage was from $2989\ \mathrm{\AA}$ to $10200 \ \mathrm{\AA}$, with integration times ranging from $2\ \mathrm{s}$ to $4360\ \mathrm{s}$. 
Additionally, we used archival data of five PMS stars obtained with Ultraviolet and Visual Echelle Spectrograph (UVES, $R \sim 40,000$) mounted on the VLT. The wavelength coverage was from $3732\ \mathrm{\AA}$ to $9496\ \mathrm{\AA}$, with integration times ranging from $10\ \mathrm{s}$ to $600\ \mathrm{s}$.

%\subsection{Data Reduction}
\subsection{Photometry}
To obtain the rotational period, we analyzed the Pre-search Data Conditioning Simple Aperture Photometry (PDCSAP) fluxes observed with Transiting Exoplanet Survey Satellite (TESS). PDCSAP fluxes are derived by aperture photometry\footnote{The details are described in the webpage of National Aeronautics and Space Administration, NASA; \url{https://heasarc.gsfc.nasa.gov/docs/tess/LightCurveFile-Object-Tutorial.html}. }.
We calculated the average of PDCSAP fluxes for each star, and the standard deviation of the flux, $\sigma_1$, then removed the flux data points greater than $3\sigma_1$ above the average. We then searched for periodicity in the signals by conducting Lomb--Scargle \citep{s82} periodogram analysis. For each object, the period of the light curve, $P$, was determined. 

We believe that the main component of the period in the TESS light curve is due to stellar spots on their surface, not other sources such as occultation by a disk or mass accretion. First, the protoplanetary disk is flared with radial distance from the central stars \citep{h16}. A central star would be hidden by the outer edge of the disk, if the inclination angle of the system is large. However such edge-on disk objects, which are observed from the direction of the disks, are rare. Second, the variable light caused by the obscuration of the disk could be in the order of years. The outer edge ($\sim 100\ \mathrm{AU}$) of the Keplerian disk orbits in hundred years. A variation of hundred years is not detected during the TESS observation of $27\ $days. Third, mass accretion is predicted to be not stable, but highly-time variable, where much of the mass was accreted in short bursts (\citealp{ke90}, \citeyear{ke94}). %They found that standard models predicted much higher accretion luminosities than observed in the CTTSs. 
Therefore, we considered the period of all PMS stars to be caused by the spots, not only the periodic group, but also quasiperiodic symmetric and burster groups. We note that those discussions are only adopted in the cause of variation period. The amplitude of the light curves should involve spot modulation, occultation by a disk, or mass accretion. 

We also reject the possibility that the PMS stars show light variation caused by pulsation. \citet{r15a} identified the A-type and early F-type stars with a period of  $\leq 0.3 \, \mathrm{days}$ as pulsators. In this study, only HIP 17695, a K0-type star, shows variable light with a period of $\leq 0.3 \, \mathrm{days}$. The measured period in this study is considered to be the rotation period caused by a spot on the surface of a PMS star.

\subsection{Spectroscopy}
We used the Image Reduction and Analysis Facility (IRAF) software package\footnote{IRAF software is distributed by the National Optical Astronomy Observatories, which are operated by the Association of Universities for Research in Astronomy, Inc., under a cooperative agreement with the National Science Foundation.} for data reduction. 
The X-shooter and UVES data had already been reduced up to the point before wavelength calibration. To improve the signal-to-noise ratio ($S/N$), we combined and averaged multiple frames taken on the same day for each object, applying weights proportional to the integration time of each frame. 
The target spectra were contaminated by Earth's atmospheric absorption. We removed telluric absorption lines using the Advanced Cerro Paranal Sky Model provided by European Southern Observatory (\citealp{n12}; \citealp{j13}). 

We subtracted the photospheric absorption components from all spectra using stellar photosphere spectral models. Generally, the photospheric absorption components of the Ca II IRT and Mg I lines are strong, particularly for K-type stars. The Mg I emission component is often obscured by the photospheric absorption, as illustrated in \figref{fig:expsub}.

\subsubsection{Determination of the atmospheric parameters ($T_{\rm eff}$ and $\log\, g$), veiling value ($r$), and line broadening}

To detect the weak chromospheric emission lines, we determined the atmospheric parameters ($T_{\rm eff}$ and $\log\, g$), the amount of veiling ($r$), and line broadening of the PMS stars in two steps. We compared the observed spectra with photospheric model spectra using the Phoenix-BT-Dusty models \citep{k19} from the POLLUX database \citep{p10}. 
Initially, we obtained $280$ models with effective temperatures ranging from $2000-\ \mathrm{K}$ to $6000\ \mathrm{K}$ (in $100\ \mathrm{K}$ steps), $\log\, g$ values from $2.0$ to $5.0\ \mathrm{dex}$ (in $0.5\ \mathrm{dex}$ steps), and a solar iron abundance (\Tabref{tab:p}). This grid encompasses the range of our targets and extends to lower gravities typical of giants.

	\begin{table}[htbp]
	\caption{Physical parameters of the model spectra}\label{tab:p}%\vspace{-0.2cm}
	\centering
	\begin{tabular}{ll}
	\hline %\rowcolor[HTML]{CDEEF0} 
	Parameter   & Range \\ \hline
	$T_{\rm eff}$  & $2000-6000\ \mathrm{K}$ with $100\ \mathrm{K}$ steps \\
	$\log \, g$ & $2.0-5.0\ \mathrm{dex}$ with $0.5\ \mathrm{dex}$ steps \\
	Abundance   & fixed on $[{\rm Fe/H}] = 0$ \\
	$v \sin i$  & $0-250\ \mathrm{km\cdot s^{-1}}$ in $26\ $ steps \\
	Veiling, $r$ &  $0.0-5.0$ in $20\ $ steps \\ \hline
	\end{tabular}
	\end{table}

For the correction of rotational broadening, we convolved the model spectra with a Gaussian kernel in 26 steps of $\sigma$, corresponding to $v\sin\, i$ values ranging from $0$ to $250\ \mathrm{km\cdot s^{-1}}$. Convolution with a Gaussian function of large $\sigma$ caused the continuum level of the model spectrum to decrease because broad absorption lines often blend with other strong lines. After convolving with the Gaussian kernel, we re-normalized the continuum for all the model spectra. 

An absorption line profile of a PMS spectrum might be obscured by continuum veiling. The amount of veiling, $r$, is defined as 
	\begin{equation}
	r = \frac{W_0}{W} -1,
	\end{equation}
The equivalent width (EQW) of an absorption line in an unveiled spectrum is denoted by $W_0$, and $W$ represents the EQW of a line in a veiled spectrum. We set the veiling parameter $r$ between $0$ and $5$ in $20$ steps. First, we added the veiling value $r$ to the normalized model spectrum. We then re-normalized the continuum to unity by dividing by $(1+r)$.

%\footnote{$S(i), (i=6)$}
We subtracted each model spectrum from the observed spectra of the $64\ $ PMS stars and evaluated the fit quality. In the first step, we selected four regions for fitting the spectra, based on the work of \citet{f17}: $6050 - 6270\ \mathrm{\AA}$ (main features: Ca I, Fe I, TiO), $7020 - 7120\ \mathrm{\AA}$ (TiO), $7600 - 7720\ \mathrm{\AA}$ (K I, Fe II), and $8160 - 8222\ \mathrm{\AA}$ (Na I, V I). According to \citet{f17}, these regions are particularly suitable for determining $T_{\rm eff}$, $\log\, g$, $v\sin\, i$, and veiling. Note that the wavelength range of $5120 - 6270\ \mathrm{\AA}$ is not available for some PMS stars observed with UVES. 
For each wavelength range, we measured the standard deviation of the count values of the residual spectrum, $S(i)$. \Figref{fig:expsub} illustrates the spectral subtraction procedure for the photospheric component in PMS stars. Many Fe I absorption lines are present in both the target and model spectra, as shown in the hatched area of \figref{fig:NaI}. Emission components of Fe I may originate in the lower chromosphere \citep{v81}. Highly active T Tauri stars exhibit narrow Fe I and Fe II emission lines at rest velocity, believed to be formed in the chromosphere \citep{hp92}, and ZAMS stars also show narrow Fe I emission lines \citep{y22}. We excluded wavelengths near strong Fe I lines from the fitting range if their $gf$ value exceeds $10^{-1.3}$. 
Here we determined the $\sigma$ value of the model with the smallest $S(i)$ in each spectral region. 
Then we determined $T_{\rm eff}$, $\log\, g$, and $r$ if the sum of squares of the standard deviation, $S(i)$, is the smallest. Note that $T_{\rm eff}$ and $\log\, g$ are one value for one object, respectively, whereas different values of $r$ are determined for the different spectral regions. 
Following \citet{f17}, in the second step, we included one or two additional blue spectral regions for objects with temperatures higher than $3500\ \mathrm{K}$. For UVES data, these regions are $4400 - 4580\ \mathrm{\AA}$ (main features: Fe I), while for X-shooter data, we added $4400 - 4580\ \mathrm{\AA}$ and $5120 - 5220\ \mathrm{\AA}$ (Mg I triplet). We then recalculated the sum of the squares of $S(i)$ and re-determined $T_{\rm eff}$, $\log\ g$, $r$, and $\sigma$.

For each object, we estimated the errors in temperature and surface gravity. We defined $T_i$ as the temperature of the model with the smallest $S(i)$. The error in temperature was calculated as the standard deviation of $T_i$, given by $\sqrt{\frac{1}{n} \sum\limits_{i=1}^n (T_i - \bar{T})^2}$. The error in surface gravity was estimated using the same procedure.

\subsubsection{Veiling measurements near the chromospheric emission lines and detection of the emission lines}
\begin{figure}[htbp]
\begin{minipage}{1.0\hsize}
\begin{center}
	\resizebox{\hsize}{!}{\includegraphics[width=12cm]{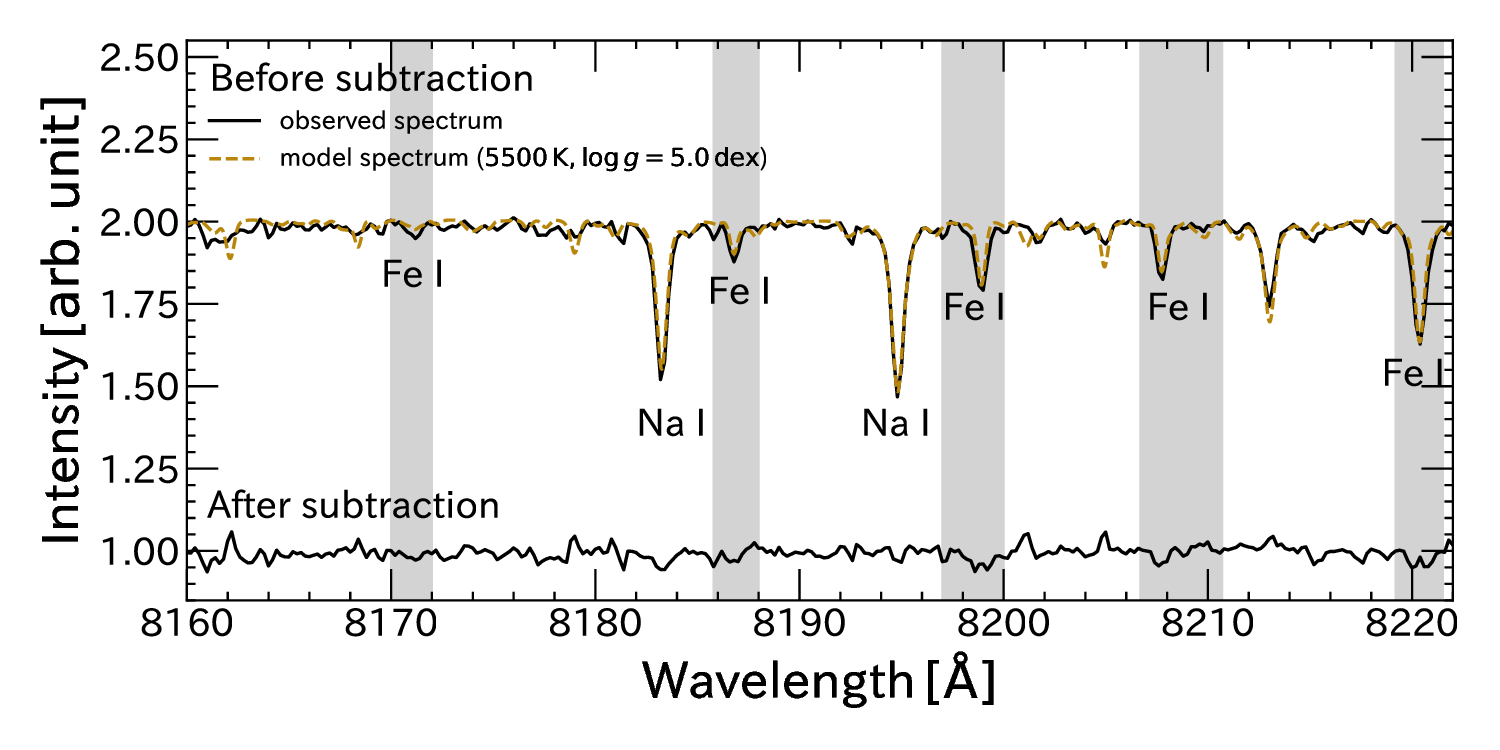}} \vspace{-0.4cm} \subcaption{Na I lines of HBC 407}\label{fig:NaI}
	\resizebox{\hsize}{!}{\includegraphics[width=12cm]{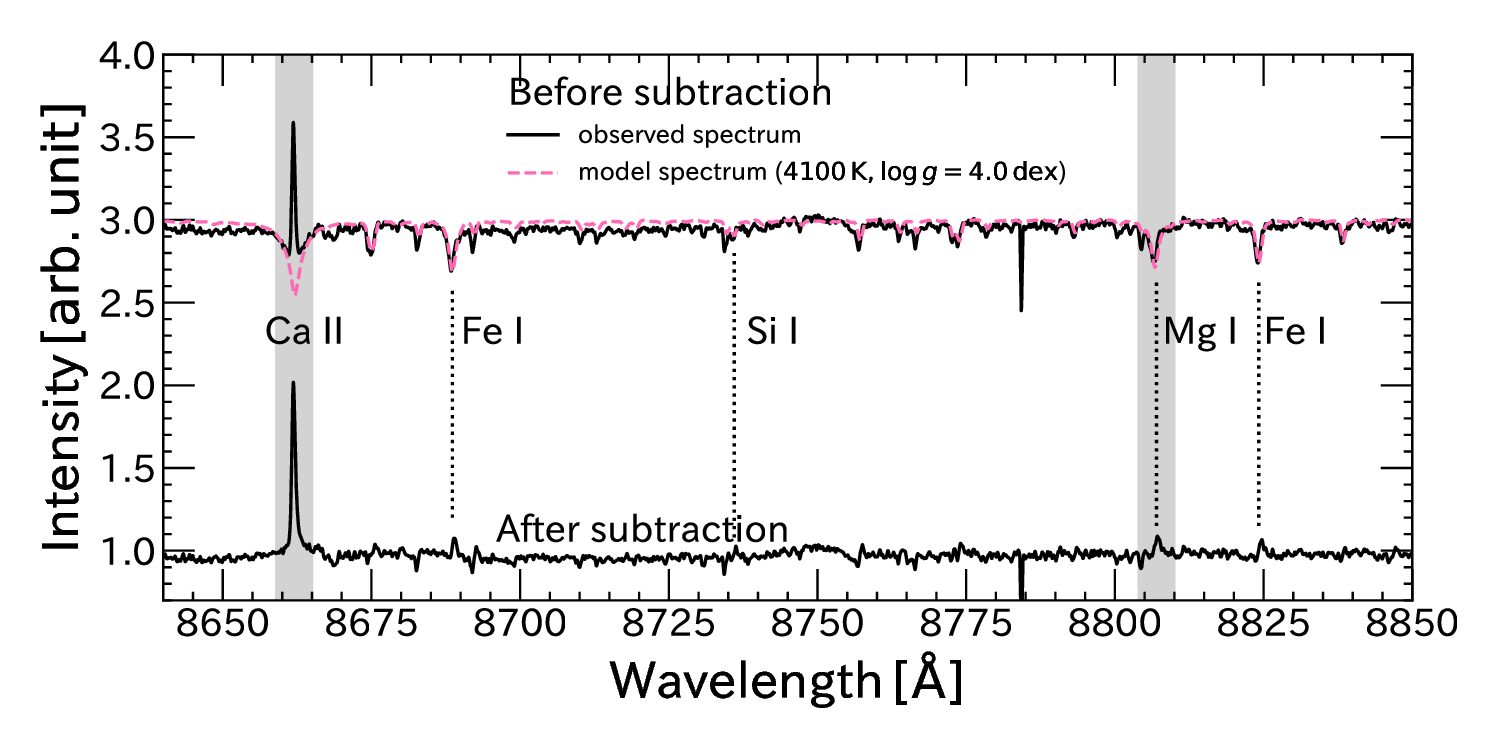}} \vspace{-0.4cm} \subcaption{Ca II and Mg I lines of AA Tau}\label{fig:recx11}
\end{center} %\label{fig:NvsR}
\end{minipage}
\caption{Procedure for subtracting the photospheric component from PMS star spectra. The top portion of each panel shows the observed spectrum (solid line), while the model spectrum is represented by the dashed line. For display purposes, the spectra of the PMS stars and the models are shifted by $+1.0$ or $+2.0$. The bottom portion of each panel displays the difference between the PMS star spectrum and the model spectrum. Hatched areas indicate Fe I, Ca II, and Mg I lines. Wavelengths within these hatched areas are excluded from the fitting range.} \label{fig:expsub}
\end{figure}

\Figref{fig:recx11} illustrates the procedure for spectral subtraction near the Ca II and Mg I lines for AA Tau. In this case, the Ca II line shows an emission profile superimposed on a broad absorption feature in the observed spectrum. We subtracted the model spectrum from the observed spectrum of the PMS star. The dashed line in \figref{fig:recx11} represents the best-fit model spectrum. We determined $r$ in the following wavelength ranges: $8420-8560\ \mathrm{\AA}$ for Ca II lines at $8498$ and $8542\ \mathrm{\AA}$, $8560-8700\ \mathrm{\AA}$ for the Ca II line at $8662\ \mathrm{\AA}$, and $8700-8850\ \mathrm{\AA}$ for the Mg I line at $8807\ \mathrm{\AA}$. Wavelengths close to the strong emission lines, shown in the hatched area in \figref{fig:recx11}, were excluded from the fitting range.

Before measuring the EQWs, we normalized the spectra after subtracting the photospheric absorption components to unity. To obtain the EQWs of the Mg I and Ca II emission lines, we integrated the areas of the corresponding emission profiles directly. The EQW errors were estimated by multiplying the standard deviation of the continuum by the wavelength range of the emission lines for each PMS star. We derived the standard deviations of the continua near the emission lines using the following wavelength ranges: $8483-8492\ \mathrm{\AA}$ for Ca II lines at $8498$ and $8542\ \mathrm{\AA}$, $8623-8632\ \mathrm{\AA}$ for Ca II at $8662\ \mathrm{\AA}$, and $8798-8802\ \mathrm{\AA}$ and $8813-8819\ \mathrm{\AA}$ for the Mg I emission line. No strong features were observed within these wavelength ranges.

\onecolumn
\twocolumn

\section{Results} \label{result}
\subsection{ $T_{\rm eff}$ and $\log \, g$, and veiling } \label{result1}

\begin{figure}[htbp]
	\centering
\resizebox{\hsize}{!}{\includegraphics[width=7cm]{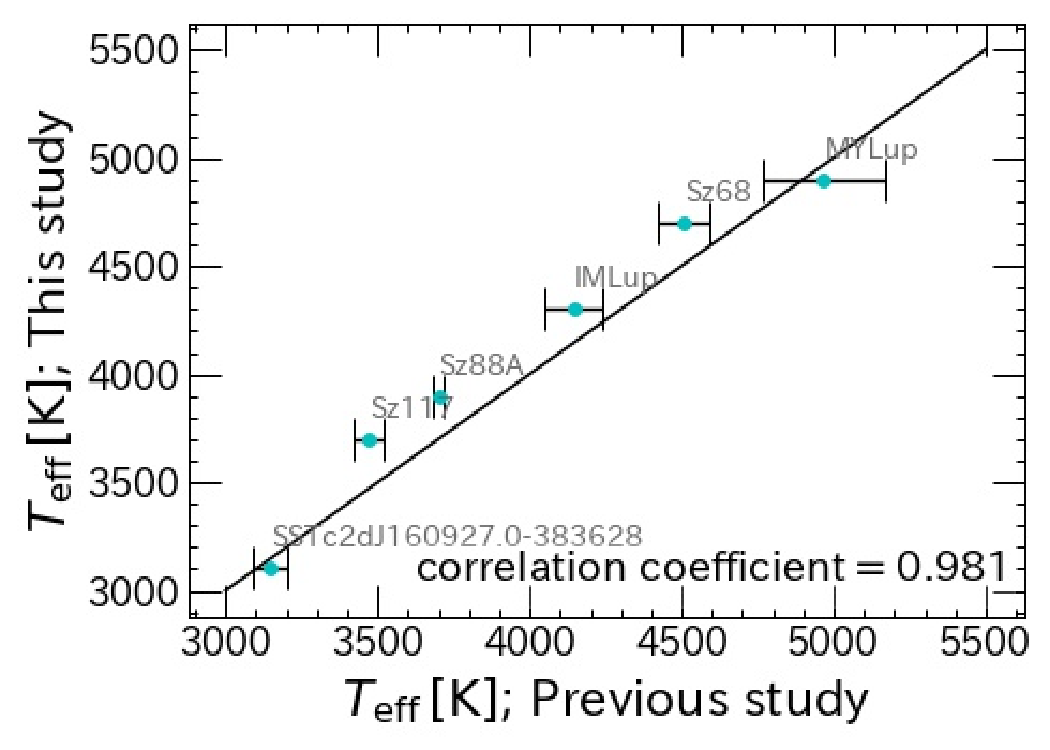}}
\caption{Comparison between the effective temperature determined in this study and that derived from the spectroscopic study of \citet{f17}. } \label{fig:TvsT}
\end{figure}

The effective temperature ($T_{\rm eff}$), surface gravity ($\log\, g$), and veiling value ($r$) for $64\ $PMS stars were determined (\tabref{tab:myTeff}). Typical uncertainties are $\approx 300\ \mathrm{K}$ for the effective temperature, and $\approx 0.8\ \mathrm{dex}$ for surface gravity. 
To assess the accuracy of $T_{\rm eff}$, we compared the $T_{\rm eff}$ values for six Lupus objects determined in this study with those reported in \citet{f17}. This comparison is shown in \figref{fig:TvsT}. The $T_{\rm eff}$ values from both studies are comparable, with a correlation coefficient of $0.981$. We conclude that the temperatures we derived for the other $64\ $PMS stars are accurate. LkCa 4 and LkCa 15 were observed on two separate occasions. The measured $T_{\rm eff}$ and $\log\,g$ values did not differ between observations, although the chromospheric emission lines exhibited time variations. This indicates that atmospheric parameters can be accurately measured even for PMS stars with time-varying chromospheric emission lines. 

	\begin{figure}[htbp]
	\centering
	\resizebox{\hsize}{!}{\includegraphics[width=8cm]{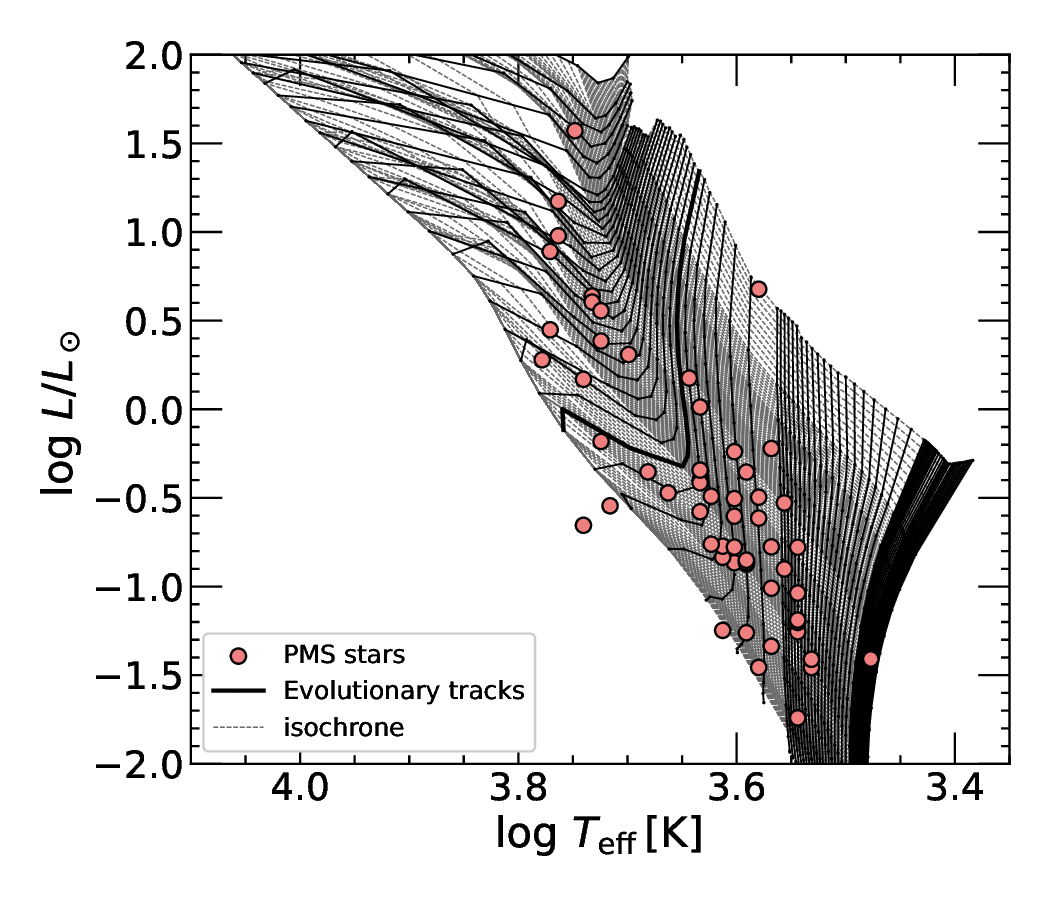}} %\vspace{-0.2cm}
	\caption{HR diagram of the investigated PMS stars. The solid lines and dashed lines denote the evolutionary tracks ($0.065 - 4.5 \ \mathrm{M_\odot}$) and isochrones ($2 \times 10^{5}\ \mathrm{yr}$ to the main-sequence) of \citet{jk07}. We linearly interpolated that model to one-tenth of the original. The thick solid line denotes the evolutionary track for a $1\ \mathrm{M_\odot}$ star. The circles represent the PMS stars. }\label{fig:hrdd}
	\end{figure}%
	
Most of the PMS stars show veiling with $r\geq 0.1$. In particular, for the highly veiled $28\ $PMS stars ($r\geq 0.6$ at $4400\ \mathrm{\AA}$ or $5120\ \mathrm{\AA}$),  amounts of veiling increases as the wavelengths decrease. In previous studies, the high veiling with a peak in the ultraviolet region has been observed for strongly accreting PMS stars. This is considered to be due to the heated photosphere below the mass accretion shock. \citet{bb90} investigated the wavelength dependence of optical veiling for $16\ $T Tauri stars. For the five T Tauri stars out of them the amounts of veiling increase with decreasing wavelengths between $4000 - 5500\ \mathrm{\AA}$ and flat for $\lambda \geq 5500\ \mathrm{\AA}$. This is similar to the results achieved this study. 

\Figref{fig:hrdd} presents the HR diagram of the investigated PMS stars. The luminosity is estimated from the $i-$band magnitude with the corrections of the interstellar extinction and the amount of the veiling. The distance of the objects was mainly taken from {\it Gaia} Data Release 2 \citep{ba18}. The distance of RY Tau and COUP 1287 was taken from {\it Gaia} Early Data Release 3 \citep{gaia21}, and that of RECX 09 was taken from \citet{m00}. We estimated the mass, age, and convective turnover time ($\tau_{\rm c}$) of the PMS stars by using the evolutionary models presented by \cite{jk07}, which is one of the few evolutionary tracks that takes into account the evolution of the internal structure of PMS stars to determine the convection turnover time. We linearly interpolated that model to one-tenth of the original. The stellar age, mass, and $\tau_{\rm c}$ are listed in \tabref{tab:objects2}.

%\input{./tables/tabB} 
%\input{./tables/tabB_mini} 
%\onecolumn
%\input{longtables/test_tabB} %\clearpage

%\input{longtables/test_tabC}
%\twocolumn

\subsection{ Chromospheric emission lines }

Mg I emission lines ($8807\ \mathrm{\AA}$) were detected from 48 out of 64 PMS stars. \Figsref{fig:result_emi2P} and \ref{fig:result_emi2Pn} show the Mg I line ($8807\ \mathrm{\AA}$) spectra of the PMS stars after subtracting the photospheric absorption. 
The EQWs and FWHMs of the chromospheric emission lines of Ca II IRT ($8542\ \mathrm{\AA}$) and Mg I ($8807\ \mathrm{\AA}$) are listed in \tabref{tab:result_of_obs2}. The PMS stars show Ca II emission lines at $8542\ \mathrm{\AA}$ whose ranges from $0.07\ \mathrm{\AA}$ to $43\ \mathrm{\AA}$. In \citet{m09}, the EQWs of the ZAMS stars are between $0.10\ \mathrm{\AA}$ and $94\  \mathrm{\AA}$. The EQWs of the stars in both studies are comparable. 
The PMS stars show Mg I emission lines whose EQWs range from $0.04\ \mathrm{\AA}$ to $0.52\ \mathrm{\AA}$. In \citet{y22}, the EQWs of the Mg I emission lines range from $0.02 \ \mathrm{\AA}$ to $0.52\ \mathrm{\AA}$ for the ZAMS stars. The EQWs of the PMS stars are comparable to those of the ZAMS stars.

%\clearpage
%\input{./tables/tabD} %\input{./tables_mini/tabDmini}
%\input{longtables/test_tabD} 
%\twocolumn

As noted by \citet{y22}, Mg I emission lines at $8807\ \mathrm{\AA}$ were detected during a total solar eclipse \citep{d68}. Additionally, some chromospheric emission lines observed during a total solar eclipse \citep{d68} are also detected in the PMS spectra after subtracting the absorption components in the $8650-8850\ \mathrm{\AA}$ range, including Si I ($8728\ \mathrm{\AA}$), P12($8750 \ \mathrm{\AA}$), and Fe I ($8680, 8824, 8838 \ \mathrm{\AA}$). Among these, P12 and Fe I at $8824\ \mathrm{\AA}$ were also observed in the T Tauri star spectra obtained by \citet{hp92}, who performed optical spectroscopy of $34\ $PMS stars with high accretion rates. \Tabref{tab:emi} provides examples of the chromospheric emission lines detected from $64$ PMS stars in this study and the number of PMS stars with detected emission lines. \citet{hp92} did not subtract the photospheric absorption components but detected Mg I and Fe I ($8807, 8824\ \mathrm{\AA}$) emission lines in eight T Tauri stars. Further, in this study, the minimum EQW of the Mg I emission lines is $0.04\ \mathrm{\AA}$, while that in \citet{hp92} was $0.37\ \mathrm{\AA}$. In this study, after subtracting the photospheric absorption components, we detected these emission lines in a larger number of PMS stars and weaker emission lines. We conclude that subtracting the model spectrum is essential for detecting weak emission lines. 

\begin{table}[htbp]
\centering
\caption{The examples of the chromospheric emission lines detected from $64\ $PMS stars }\label{tab:emi}
\begin{tabular}{lccc}
\hline
\multicolumn{2}{l}{Wavelength $\ \mathrm{[\AA]}$} &  \multicolumn{2}{c}{Number of PMS stars} \\ 
{} & {}  & \citet{hp92} & This work \\ \hline%
%Pa18&8440&25\\
%O I & 8446 & 20 \\
Si I         & 8728     & -        & 15/64 \\
Pa12         & 8750   & 24/34      & 23/64  \\
Mg I         & 8807    &8/34      & 47/64 \\
Fe I         & 8824     &  8/34     & 41/64 \\ 
Fe I         & 8838     & -       & 27/64 \\ \hline
\end{tabular}
\end{table}

\section{Discussion} \label{discussion}
\subsection{Veiling and infrared excess}\label{sec:4.1.3}

\begin{figure}[htbp]
\centering
   \resizebox{\hsize}{!}{\includegraphics[width=9cm]{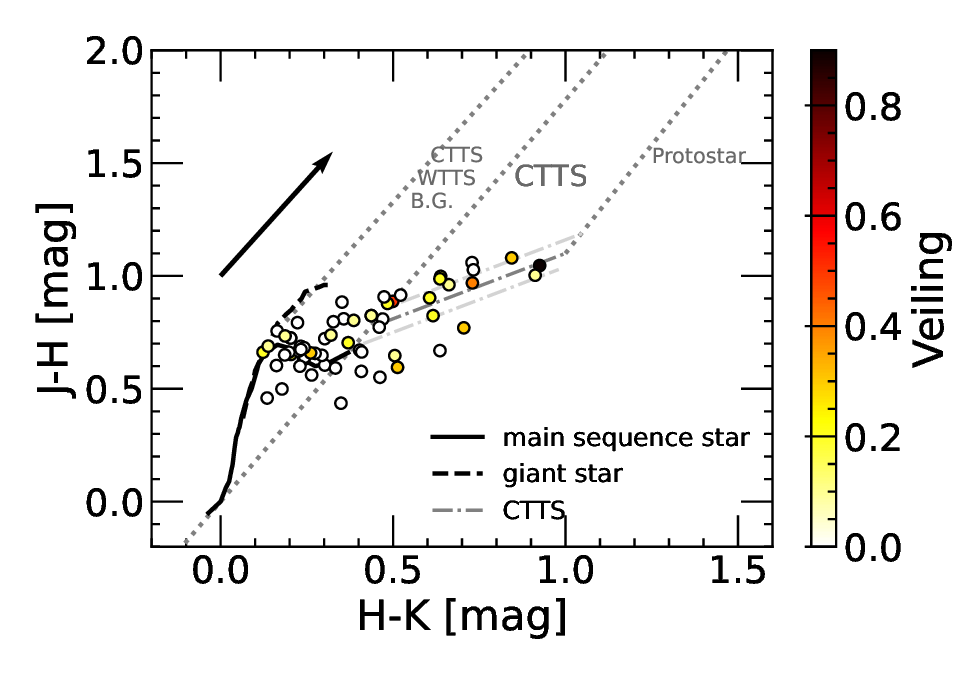}} \vspace{-0.6cm}
\caption{Near-infrared color-color diagram for the PMS stars we investigated. Circles denote the PMS stars, with the color indicating the amount of veiling measured in $8798-8802$ and $8813-8819\ \mathrm{\AA}$. The dash-dot lines represent the linear least-squares fit to the intrinsic color of the CTTSs \citep{m97}. Intrinsic colors for main-sequence stars (solid line) and giant stars (dashed line) are from \citet{b88}. B.G. stands for background star. The arrow and dotted lines show the interstellar reddening vector from \citet{c81}. }\label{fig:cc}
\end{figure}

\Figref{fig:cc} is the near-infrared color-color diagram for the PMS stars. We estimated the quantities of intrinsic near-infrared excess using the method described in \citet{h19}. The amount of near-infrared excess for the PMS stars is determined by the intersection of the reddening vector originating from the observed $JHK$ colors and the intrinsic colors for CTTSs. The zero point of the intrinsic near-infrared excess is defined as the intersection of a line parallel to the reddening vector through an M6 dwarf color and the intrinsic CTTS line, $(J-H, H-K) = (0.79, 0.47)$. The point of unity for near-infrared excess is defined as the reddest intrinsic color of CTTSs, $(J-H, H-K) = (1.1, 1.0)$ \citep{m97}. Excess values are listed in \tabref{tab:objects1}, ranging from $-0.68$ to $0.91$. Out of the PMS stars in this study, $37$ show no excess ($\leq 0$), while $14\ $PMS stars show excess ($>0$). We did not calculate excess values for PMS stars below the line parallel to the reddening vector through an M6 dwarf color and the intrinsic CTTS line. 

In section \ref{result1}, we reported that some highly veiled PMS stars exhibit a greater amount of veiling at shorter wavelengths, which is thought to be due to the heated photosphere below the shock resulting from mass accretion from a protoplanetary disk. The color-color diagram is a valuable tool for examining the presence of protoplanetary disks. However, one-third of CTTSs do not show significant infrared excess \citep{m97}. Spectroscopy allows us to detect veiling caused by the heated photosphere, indicating mass accretion from a protoplanetary disk. Indeed, the six PMS stars without excess in \figref{fig:cc} show veiling amounts of $0.1-0.2$ in the $8798-8802$ and $8813-8819\ \mathrm{\AA}$ regions.

\subsection{ Chromospheric emission lines }

In this section, we investigate the strength of the Ca II and Mg I emission lines to determine whether the chromosphere of the PMS stars is activated by the dynamo process or mass accretion. When discussing dynamo activity, we consider not only rotation, but also the convective turnover time, $\tau_{\rm c}$. We calculated the Rossby number, $N_{\rm R}$, as follows: 
	\begin{equation}
	\label{eq:nr}
	N_{\rm R} = \frac{P}{\tau_{\rm c}} = \frac{2 \pi R}{\tau_{\rm c} v\sin i}.
	\end{equation}
For $45\ $PMS stars, we used the rotation period $P$ derived from TESS light curves. Details of the data reduction will be discussed in a subsequent paper. Unfortunately, for $19\ $PMS stars not observed with TESS, we obtained $v \sin i$ from the Catalog of Stellar Rotational Velocities \citep{g05}. 

The stellar radius $R$ was estimated from the $i-$band magnitude and the effective temperature using Stefan-Boltzmann's law. We used the $i$-band magnitudes (in the AB system) from the Fourth United States Naval Observatory CCD Astrograph Catalog (UCAC4 Catalog, \cite{za13}) and the ATLAS All-Sky Stellar Reference Catalog \citep{t18}, accounting for interstellar extinction ($A_i$) and veiling. 
To obtain the veiling amount ($r$) in the wavelength range correspond to $i-$band, the amounts of the veiling in $7020$, $7600$, and $8160\ \mathrm{\AA}$ were averaged. 

We estimate $\tau_{\rm c}$ of the PMS stars with evolutionary tracks presented in \citet{jk07}, as shown in Section \ref{result1} and \figref{fig:hrdd}. The rotational period, $\tau_{\rm c}$, and Rossby numbers are listed in \tabref{tab:objects2}.  

We calculate the strengths of the emission lines, $F^{\prime}$, for the Ca II line at $8542\ \mathrm{\AA}$ and Mg I line at $8807\ \mathrm{\AA}$. A bolometric continuum flux per unit area at a stellar surface, $F$, was calculated at first. $F$ is given as
	\begin{equation} 
	\log \frac{f}{f_0} = - \frac{2}{5} \left(m_{i, {\rm AB}} - A_i \right).
	\end{equation}
where $f$ is the bolometric continuum flux of the object per unit area as observed on the Earth. The bolometric continuum flux per unit area for an $m_i = 0\ \mathrm{mag}$ (the AB system) star, $f_0$, is $1.852 \times 10^{-12}\ \mathrm{W \cdot m^{-2} \cdot \AA^{-1}}$ (\citealp{f96}). For an object without veiling, $F$ is written as 
	\begin{equation} 
	F = f \cdot \left(\frac{d}{R} \right)^2,
	\end{equation}
where $d$ denotes the distance to the object. For an object with veiling, $F$ is modified as 
	\begin{equation} 
	F = \frac{f}{1+r} \cdot \left(\frac{d}{R} \right)^2, 
	\label{eq4f}
	\end{equation}

$F^{\prime}$ is derived by multiplying $F$ by the EQW of the emission lines, $W$. 
For an object without veiling, $F^{\prime}$ is written as
	\begin{equation} 
	F^{\prime} = F \cdot W_0.
	\end{equation}
Thus, for an object with veiling, 
	\begin{eqnarray}
	F^{\prime} & = & F \cdot W_0 \\ \nonumber
	{}         & = & F \cdot (1+r) W %\\ \nonumber
%	{}         & = & \left(\frac{f}{1+r}\right) \left(\frac{d}{R} \right)^2 \cdot (1+r) W\\ \nonumber
%	{}         & = & f \left(\frac{d}{R} \right)^2 \cdot W
	\end{eqnarray} %$r$ is the veiling value measured near each of the emission lines.
With $F^{\prime}$ and $T_{\rm eff}$, we calculated $R^{\prime}$ for the PMS stars as
	\begin{equation}
	R^{\prime} = \frac{F^{\prime}}{\sigma T_{\rm eff}^4}, 
	\label{eq4}
	\end{equation}
where $\sigma$ is Stefan-Boltzmann's constant. The $R^{\prime}$ values for the chromospheric emission lines of  Ca II IRT ($8542 \ \mathrm{\AA}$) and Mg I ($8807\ \mathrm{\AA}$) are listed in \tabref{tab:result_of_obs2}.

\subsubsection{ Chromospheric Ca II emission lines } \label{ssc2}

\begin{figure}[htbp]
\centering
   \resizebox{\hsize}{!}{\includegraphics[width=9cm]{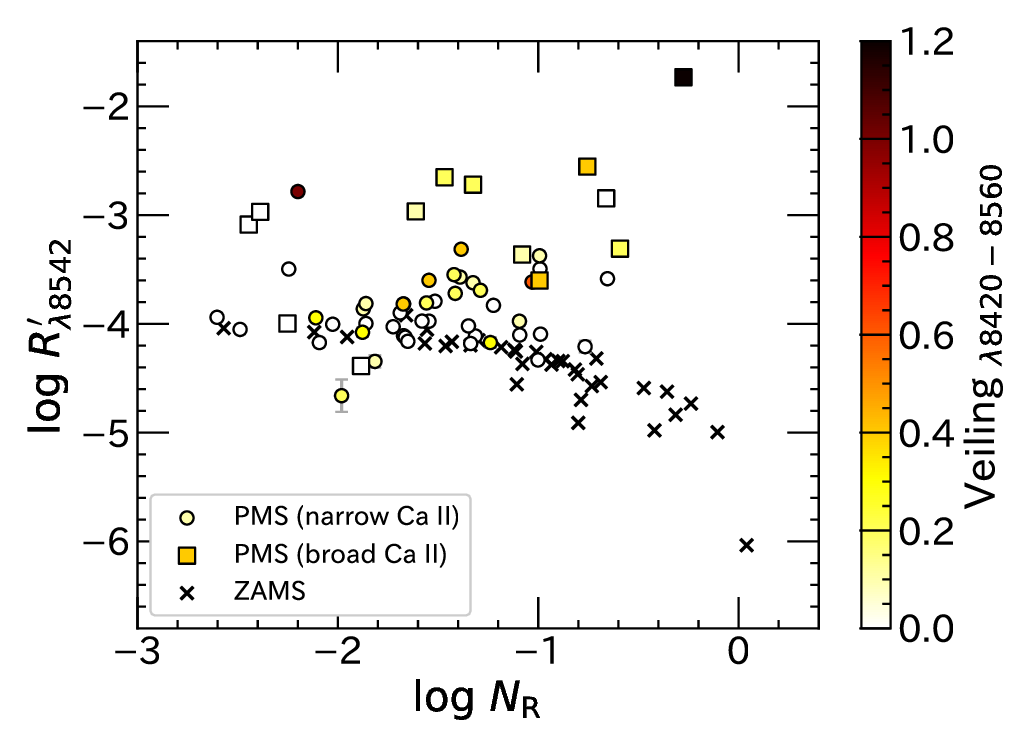}} \vspace{-0.2cm}
\caption{Relationship between the Rossby number, $N_{\rm R}$, and the ratio of the surface flux of the chromospheric Ca II IRT emission line at $8542\ \mathrm{\AA}$ to the stellar bolometric luminosity, $R^{\prime}$. Circles represent PMS stars with narrow Ca II emission lines ($\leq 100\ \mathrm{km\cdot s^{-1}}$), while squares represent PMS stars with broad Ca II emission lines ($> 100\ \mathrm{km\cdot s^{-1}}$). Crosses denote ZAMS stars (\citealp{m09}; \citealp{y22}). The color of the circles indicates the amount of veiling measured in $8420-8560\ \mathrm{\AA}$. }\label{fig:40}
\end{figure}

\Figref{fig:40} shows $R^{\prime}$ of the Ca II IRT emission line ($8542\ \mathrm{\AA}$) as a function of the Rossby number. The color of the symbols represents the veiling value measured in $8420 - 8560 \ \mathrm{\AA}$. 
Most PMS stars without veiling have $N_{\rm R}$ and $R^{\prime}$ values similar to those of ZAMS stars in the saturated regime. They also exhibit narrow Ca II emission lines with FWHMs $\leq 100\ \mathrm{km\cdot s^{-1}}$. We propose that the chromospheres of these PMS stars are activated by the dynamo process and are completely filled by the emitting region. This is consistent with previous studies indicating that PMS stars have optically thick Ca II IRT emission lines (e.g., \citealp{BB}; \citealp{hp92}). 

Some PMS stars show larger $R^{\prime}$ values than ZAMS stars with the same Rossby numbers. Most of these PMS stars have veiling ($r \geq 0.1$), and a portion of them exhibit broad Ca II emission lines with widths $> 100\ \mathrm{km\cdot s^{-1}}$. These broad Ca II IRT emission features in PMS stars are thought to originate from active chromospheres heated by accretion.

\subsubsection{Chromospheric Mg I emission lines} \label{sec:chMg}

\begin{figure}[hbtp]
\centering
   \resizebox{\hsize}{!}{\includegraphics[width=9cm]{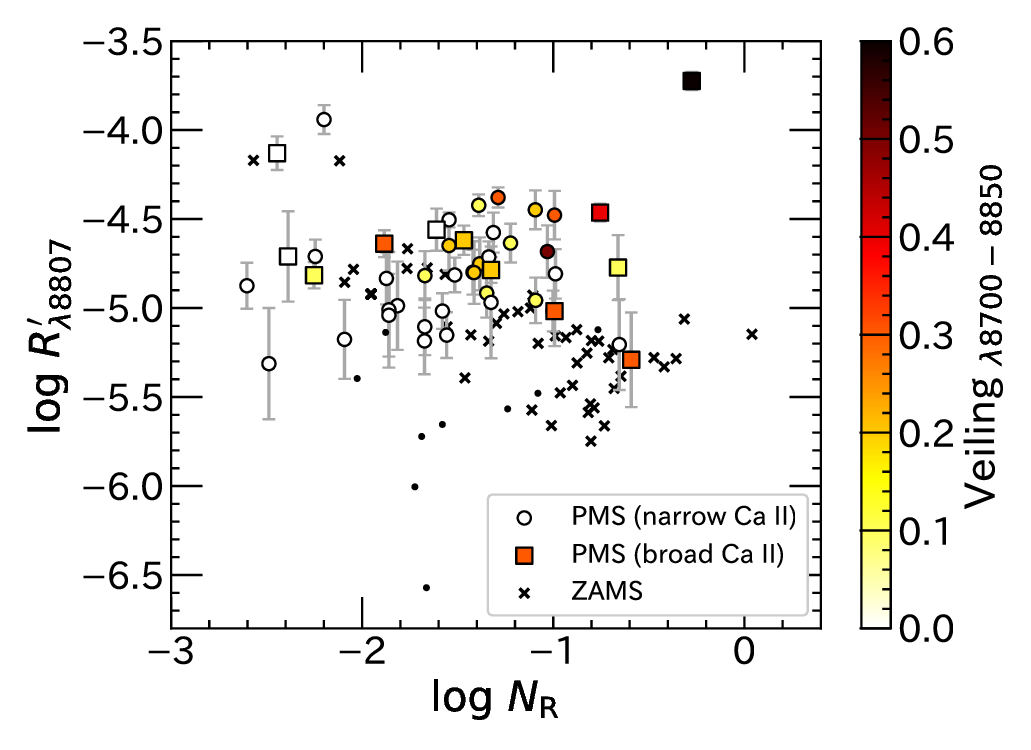}} \vspace{-0.6cm}
\caption{Relationship between the ratio of the surface flux of the chromospheric Mg I emission line to the stellar bolometric luminosity, $R^{\prime}$, and the Rossby number, $N_{\rm R}$. Large circles represent PMS stars with narrow Ca II emission lines ($\leq 100\ \mathrm{km\cdot s^{-1}}$), while squares represent PMS stars with broad Ca II emission lines ($> 100\ \mathrm{km\cdot s^{-1}}$). The color of the large circles and squares indicates the amount of veiling measured in $8700 - 8850\ \mathrm{\AA}$. Dots represent PMS stars with no significant detection of the Mg I emission line. ZAMS stars examined in \citet{y22} are shown as cross symbols. }\label{fig:MgIeV}
\end{figure}

\Figref{fig:MgIeV} shows the ratio of the surface flux of the Mg I emission line at $8807\ \mathrm{\AA}$ to the stellar bolometric luminosity, $R^{\prime}$, as a function of the Rossby number. PMS stars are compared with ZAMS stars examined in \citet{y22}, and are colored to represent the amount of veiling measured in $8700 - 8850\ \mathrm{\AA}$. 

%Compared to the ZAMS stars 
With the same Rossby number, some PMS stars have $R^{\prime}$ values comparable to those of ZAMS stars, while others exhibit larger $R^{\prime}$ values. PMS stars without veiling show emission line strengths similar to those of ZAMS stars with the same Rossby numbers. The Mg I emission lines of such stars are considered to be formed by a dynamo process similar to that in ZAMS stars. 

On the other hand, PMS stars with larger $R^{\prime}$ than ZAMS stars with the same Rossby number, such as CV Cha and TWA 01, show veiling. More than half of these stars, including DK Tau, GQ Lup, and IQ Tau, also exhibit near-infrared excess as measured in section \ref{sec:4.1.3} and broad Ca II emission lines with $> 100\ \mathrm{km\cdot s^{-1}}$. These objects are thought to possess protoplanetary disks. For such stars, the chromosphere is activated not only by the dynamo process but also by mass accretion from a protoplanetary disk. 

Veiling originates in the accretion shock column, which is generally divided into three subregions: the preshock region, the postshock (cooling) region, and the heated photosphere (\citealp{cm98}; \citealp{h16}). 
The stellar photospheric pressure at the depth of continuum formation is calculated as $\log\ g\approx 3.5 \sim 10^4\ \mathrm{dyn\cdot cm^{-2}}$, indicating that the shock forms near the photosphere \citep{cm98}. The Mg I emission lines are known to form from the lower chromosphere to the upper photosphere \citep{f94}, and it is considered that these emission lines are induced by accretion because the height where Mg I forms aligns with the shock region. 

With the completion of this study, we propose using Mg I chromospheric emission lines as a new and effective method for examining the activity of fast-rotating PMS and ZAMS stars. \citet{noyes} identified five observational indicators of the dynamo process: rotation, stellar mass, spectral type, convective zone thickness, and convective turnover time. They suggested that Rossby numbers, which incorporate these parameters, are a useful indicator of stellar activity for main-sequence stars. %Linsky is ...

We introduce mass accretion as an additional mechanism influencing chromospheric activity in PMS stars. \citet{y20} studied Ca II emission lines in PMS stars and suggested that those with high mass accretion rates exhibit stronger Ca II emission lines than ZAMS stars. However, this conclusion was not strongly supported due to saturation of $R^{\prime}$ against the Rossby numbers. \citet{y22} demonstrated that Mg I emission lines are a good indicator of activity in fast-rotating and active ZAMS stars. We conclude that Mg I emission lines are more suitable than Ca II emission lines for assessing chromospheric activity in active objects such as PMS stars.

%\onecolumn

%\twocolumn

\section{Conclusions}
We investigated the chromospheric activity of $64\ $G, K, and M-type PMS stars in four star-forming regions and seven moving groups. High-resolution optical spectra of these PMS stars were obtained using VLT/X-shooter and VLT/UVES, and the infrared Mg I ($8807\ \mathrm{\AA}$) emission lines were examined. 
To detect the weak chromospheric emission lines, we first determined the atmospheric parameters ($T_{\rm eff}$ and $\log \, g$), projected rotational velocity, and veiling of each PMS star by comparing the observed spectra with photospheric model spectra. After subtracting the photospheric model spectrum from the PMS spectrum, Mg I emission lines were detected. 
For PMS stars with no veiling, the strengths of the Mg I emission lines are comparable to those of ZAMS stars with similar Rossby numbers. These lines are believed to form through a dynamo process similar to that in ZAMS stars. 
In contrast, PMS stars with veiling exhibit stronger Mg I emission lines than ZAMS stars. These stars are likely surrounded by protoplanetary disks, where mass accretion generates shocks near the photosphere, heating the chromosphere. We propose that the chromosphere of PMS stars is activated not only by the dynamo process but also by mass accretion.

\begin{acknowledgements}
This work would not have been possible without the financial support of the Japan Society for the Promotion of Science (JSPS) KAKENHI grant number 23KJ1855. M.Y. is also supported as a JSPS Research Fellow (DC2 and PD). 

We would like to thank Prof. Jung and Prof. Kim for providing the numerical results of the evolution tracks from \citet{jk07} through private communication (see \figref{fig:hrdd}). This paper includes data collected with the TESS mission, from the Mikulski Archive for Space Telescopes (MAST) data archive at the Space Telescope Science Institute (STScI).
\end{acknowledgements}

\begin{appendix}
\section{Figures and tables}
\onecolumn
\begin{longtable}[H]{lrrrrrrcl}
\caption{\label{tab:objects1} Physical parameters of $64\ $PMS stars.}\\
\hline\hline
Object Name             & $i$    & $J$    & $H$    & $K$         & $A_V$        & Excess & Distance     & Instrument \\
                        &  &  &  &   &   &   & $\mathrm{[pc]}$ &            \\
(1)                     & (2)       & (3)       & (4)       & (5)       & (6)       & (7)        & (8)      & (9)        \\    % table heading
\hline
\endfirsthead
\caption{continued.}\\
\hline\hline
Object Name             & $i$    & $J$    & $H$    & $K$         & $A_V$        & Excess & Distance     & Instrument \\
                        &  &  &  &   &   &   & $\mathrm{[pc]}$ &            \\
(1)                     & (2)       & (3)       & (4)       & (5)       & (6)       & (7)        & (8)      & (9)        \\
\hline
\endhead
\hline
\endfoot
\hline
\multicolumn{2}{@{}l@{}}{\hbox to0pt{\parbox{158mm}{\small
References for parameters: (2) $i-$mag: UCAC4 Catalog \citep{za13}. (3)(4)(5) $J-$mag, $H-$mag and $K_{\rm s}-$mag: 2MASS survey \citep{c03}. (6) Interstellar extinction at $V-$band: \citealp{he14}, \citealp{wa10}, \citealp{go18}, \citealp{yu23}, \citealp{f09}, \citealp{y12}, \citealp{d12}, \citealp{m21}. (8) Distance: {\it Gaia} DR2 \citep{ba18}.
}\hss}} 
\endlastfoot
\multicolumn{3}{l}{Taurus molecular cloud}      &           &           &           &            &          &            \\ \hline
AA Tau                  & 12.73     & 9.43      & 8.55      & 8.05      & 0.40      & 0.00       & 137      & UVES       \\
BP Tau                  & 10.98     & 9.10      & 8.22      & 7.74      & 0.45      & -0.02      & 129      & UVES       \\
CQ Tau                  & 9.57      & 7.93      & 7.06      & 6.17      & 0.40      &            & 162      & X-shooter  \\
DE Tau                  & 11.19     & 9.18      & 8.27      & 7.80      & 0.35      & -0.09      & 127      & X-shooter  \\
DF Tau                  & 10.35     & 8.17      & 7.26      & 6.73      & 0.10      & 0.02       & 125      & UVES       \\
DG Tau                  & 11.47     & 8.69      & 7.72      & 6.99      & 1.60      & 0.49       & 121      & X-shooter  \\
DK Tau                  & 11.34     & 8.72      & 7.76      & 7.10      & 0.70      & 0.32       & 128      & X-shooter  \\
DM Tau                  & 12.57     & 10.44     & 9.76      & 9.52      & 0.10      & -0.39      & 145      & X-shooter  \\
DN Tau                  & 10.95     & 9.14      & 8.34      & 8.02      & 0.55      & -0.32      & 128      & X-shooter  \\
DR Tau                  & 10.83     & 8.85      & 7.80      & 6.87      & 0.45      & 0.88       & 195      & UVES       \\
FP Tau                  & 12.04     & 9.90      & 9.18      & 8.87      & 0.60      & -0.27      & 128      & X-shooter  \\
GM Aur                  & 11.17     & 9.34      & 8.60      & 8.28      & 0.30      & -0.25      & 159      & X-shooter  \\
HBC 407                 & 11.93     & 10.58     & 10.08     & 9.90      & 0.80      & -0.27      & 126      & X-shooter  \\
HP Tau                  & 12.36     & 9.55      & 8.47      & 7.63      & 3.15      & 0.62       & 176      & X-shooter  \\
IQ Tau                  & 12.35     & 9.42      & 8.42      & 7.78      & 0.85      & 0.20       & 131      & X-shooter  \\
LkCa 04                 & 11.34     & 9.25      & 8.52      & 8.32      & 0.35      & -0.53      & 129      & X-shooter  \\
LkCa 15                 & 10.97     & 9.42      & 8.60      & 8.16      & 0.30      & -0.07      & 158      & X-shooter  \\
RY Tau                  & 9.30      & 7.16      & 6.13      & 5.40      & 1.85      & 0.41       & 444      & X-shooter  \\
T Tau                   & 9.10      & 7.24      & 6.24      & 5.33      & 1.25      & 0.91       & 144      & X-shooter  \\
UX Tau                  & 9.98      & 8.62      & 7.96      & 7.55      & 0.65      &            & 139      & X-shooter  \\
V1354 Tau               & 10.47     & 9.28      & 8.82      & 8.69      & 0.40      & -0.32      & 122      & X-shooter  \\
                        &           &           &           &           &           &            &          &            \\ \hline
\multicolumn{3}{l}{${\rm\rho}$ Ophiuchi molecular cloud} &           &           &           &            &          &            \\ \hline
V2062 Oph               & 11.21     & 9.23      & 8.25      & 7.61      & 3.80      & 0.21       & 145      & X-shooter  \\
V2129 Oph               & 10.48     & 8.44      & 7.67      & 7.21      & 0.25      & 0.07       & 130      & X-shooter  \\
V2247 Oph               & 11.53     & 9.42      & 8.63      & 8.41      & 7.10      & -0.58      & 112      & X-shooter  \\
                        &           &           &           &           &           &            &          &            \\ \hline
\multicolumn{3}{l}{Lupus star-forming region}  &           &           &           &            &          &            \\ \hline
GQ Lup                  & 10.58     & 8.61      & 7.70      & 7.10      & 1.60      & 0.26       & 151      & X-shooter  \\
                        &           &           &           &           &           &            &          &            \\ \hline
\multicolumn{3}{l}{Chamaeleon I star-forming region} &           &           &           &            &          &            \\ \hline
CHX22                   & 10.24     & 8.65      & 8.04      & 7.88      & 1.80      & -0.46      & 203      & X-shooter  \\
CR Cha                  & 10.07     & 8.46      & 7.82      & 7.31      & 0.56      &            & 186      & X-shooter  \\
CV Cha                  & 10.00     & 8.29      & 7.46      & 6.85      & 0.62      & 0.40       & 192      & X-shooter  \\
CW Cha                  & 13.64     & 10.91     & 9.85      & 9.13      &           & 0.35       & 195      & X-shooter  \\
                        &           &           &           &           &           &            &          &            \\ \hline
\multicolumn{3}{l}{Orionis OB 1c association} &           &           &           &            &          &            \\ \hline
CO Ori                  & 10.18     & 7.98      & 7.21      & 6.51      & 2.00      &            & 400      & X-shooter  \\
COUP 1287               & 16.58     & 14.89     & 14.31     & 13.90     & 1.13      &            & 585      & X-shooter  \\
COUP 1423               & 13.91     & 11.78     & 10.97     & 10.50     & 0.62      & 0.04       & 413      & X-shooter  \\
GW Ori                  & 9.02      & 7.70      & 7.10      & 6.59      & 1.30      &            & 398      & X-shooter  \\
HBC 167                 & 10.38     & 9.43      & 9.00      & 8.65      & 0.40      &            & 407      & X-shooter  \\
OV Ori                  & 13.67     & 11.51     & 10.70     & 10.32     & 1.04      & -0.17      & 390      & X-shooter  \\
V1044 Ori               & 10.91     & 9.70      & 9.15      & 8.69      & 2.00      &            & 388      & X-shooter  \\
V1458 Ori               & 13.28     & 10.85     & 9.97      & 9.62      & 0.59      & -0.38      & 1211     & X-shooter  \\
                        &           &           &           &           &           &            &          &            \\ \hline
\multicolumn{3}{l}{Upper Scorpius association} &           &           &           &            &          &            \\ \hline
2MASS J16123916-1859284 & 12.43     & 10.28     & 9.47      & 9.11      & 0.57      & -0.26      & 139      & X-shooter  \\
V1149 Sco               & 9.66      & 8.36      & 7.69      & 7.05      & 0.45      &            & 165      & X-shooter  \\
                        &           &           &           &           &           &            &          &            \\ \hline
\multicolumn{3}{l}{AB Doradus moving group} &           &           &           &            &          &            \\ \hline
HIP 17695               & 9.78      & 7.80      & 7.17      & 6.93      & 0.06      & -0.30      & 17       & X-shooter  \\
                        &           &           &           &           &           &            &          &            \\ \hline
\multicolumn{3}{l}{${\rm\beta}$ Picoris moving group} &           &           &           &            &          &            \\ \hline
AU Mic                  & 7.36      & 5.44      & 4.83      & 4.53      & -0.90     & -0.10      & 10       & UVES \\
                        &           &           &           &           &           &            &          &            \\ \hline
\multicolumn{3}{l}{"Cha-Near" region} &           &           &           &            &          &            \\ \hline
RX J1147.7-7842         & 11.56     & 9.52      & 8.86      & 8.59      &           & -0.25      & 101      & X-shooter  \\
RX J1204.6-7731         & 11.80     & 9.77      & 9.13      & 8.88      & 0.00      & -0.30      & 100      & X-shooter  \\
                        &           &           &           &           &           &            &          &            \\ \hline
\multicolumn{3}{l}{${\rm\eta}$ Chamaeleontis cluster} &           &           &           &            &          &            \\ \hline
RECX 04                 & 11.41     & 9.54      & 8.78      & 8.62      &           & -0.68      & 99       & X-shooter  \\
RECX 06                 & 12.36     & 10.23     & 9.58      & 9.29      &           & -0.18      & 98       & X-shooter  \\
RECX 07                 & 10.03     & 8.42      & 7.76      & 7.64      &           & -0.65      & 98       & X-shooter  \\
RECX 09                 & 12.76     & 10.26     & 9.67      & 9.34      & 0.00      &            & 1547     & X-shooter  \\
RECX 10                 & 11.31     & 9.65      & 8.92      & 8.73      &           & -0.59      & 98       & X-shooter  \\
RECX 11                 & 10.32     & 8.73      & 8.03      & 7.66      & 0.00      & -0.07      & 98       & X-shooter  \\
RECX 15                 & 12.67     & 10.51     & 9.83      & 9.43      & 0.00      &            & 92       & X-shooter  \\
                        &           &           &           &           &           &            &          &            \\ \hline
\multicolumn{3}{l}{TW Hydrae association}           &           &           &           &            &          &            \\ \hline
TWA 1                  & 9.99      & 8.22      & 7.56      & 7.30      & 0.00      & -0.29      & 60       & X-shooter  \\
TWA 5B                  & 9.71      & 7.67      & 6.99      & 6.75      &           & -0.37      & 49       & X-shooter  \\
TWA 6                  & 10.28     & 8.87      & 8.18      & 8.04      & 0.05      & -0.65      & 66       & X-shooter  \\
TWA 7                  & 9.97      & 7.79      & 7.13      & 6.90      & -0.10     & -0.39      & 34       & X-shooter  \\
TWA 8A                  & 10.43     & 8.34      & 7.66      & 7.43      & 0.05      & -0.38      & 46       & X-shooter  \\
TWA 8B                  & 13.08     & 9.84      & 9.28      & 9.01      & 0.20      & -0.13      & 46       & X-shooter  \\
TWA 9A                  & 10.15     & 8.68      & 8.03      & 7.85      & -0.05     & -0.47      & 76       & X-shooter  \\
TWA 9B                  & 12.01     & 9.98      & 9.38      & 9.15      & 0.00      & -0.28      & 76       & X-shooter  \\
TWA 14                  & 11.26     & 9.42      & 8.73      & 8.50      & 0.10      & -0.40      & 92       & X-shooter  \\
TWA 22                  & 11.26     & 8.55      & 8.09      & 7.69      &           &            & 20       & X-shooter  \\
TWA 25                  & 9.86      & 8.17      & 7.50      & 7.31      & 0.05      & -0.46      & 53       & X-shooter  \\
\end{longtable}

%\clearpage
\begin{table}[htbp]
  \centering
  \caption{Details of the archive data of VLT/X-shooter and VLT/UVES}
\begin{tabular}{llc}
\hline
Program ID    & Observation Date                   & Integration \\ 
              &                                    & Time$\ \mathrm{[s]}$\\ \hline
\multicolumn{3}{l}{VLT / X-shooter}                                           \\ \hline
084.C-1095(A) & 2010-01-18, 19, 20                 & 2 - 1000                 \\
085.C-0238(A) & 2010-04-05, 06, 07                 & 20 - 3600                \\
085.C-0764(A) & 2010-05-04                         & 480 - 1800               \\
086.C-0173(A) & 2011-01-12, 13                     & 100 - 400                \\
088.C-0218(A) & 2011-11-05                         & 20 - 200                 \\
088.D-0556(A) & 2012-02-05                         & 48 - 96                  \\
288.C-5013(A) & 2012-02-27, 2012-03-06             & 800 - 880                \\
090.C-0050(A) & 2012-11-13, 15, 24                 & 10 - 1200                \\
088.C-0218(E) & 2012-12-17                         & 8 - 60                   \\
090.C-0253(A) & 2013-03-14                         & 1200 - 1440              \\
093.C-0506(A) & 2014-05-06                         & 480 - 880                \\
093.C-0757(A) & 2014-06-04                         & 300 - 600                \\
094.C-0913(A) & 2014-12-05, 08, 09, 10, & 30 - 1720                \\
              & 2015-01-21 &                 \\
094.C-0805(A) & 2015-01-15, 2015-02-19             & 25 - 388                 \\
094.C-0327(A) & 2015-01-16                         & 10 - 240                 \\
095.C-0147(A) & 2015-05-01                         & 3400 - 3840              \\
095.D-0949(A) & 2015-06-12                         & 30 - 120                 \\
096.C-0455(A) & 2015-10-23                         & 60 - 65                  \\
096.C-0979(A) & 2015-10-24, 28                     & 30 - 360                 \\
097.C-0669(A) & 2016-05-03                         & 100 - 4360               \\
097.C-0378(A) & 2016-07-25                         & 30 - 480                 \\
101.C-0389(B) & 2018-06-16                         & 2700 - 3000              \\
101.C-0866(A) & 2018-05-19, 20                     & 30 - 800                 \\
103.C-0887(B) & 2019-06-24                         & 20 - 300                 \\
106.20Z8.004  & 2021-05-02, 03                     & 10 - 300                 \\
108.2206.001  & 2021-10-13, 22,                    & 60 - 2800                \\
              & 2021-11-18, 21                     &                          \\
106.20Z8.006  & 2021-11-24, 26, 2021-12-02         & 30 - 800                 \\
106.20Z8.008  & 2022-11-30                         & 80 - 400    \\ \hline 
\multicolumn{3}{l}{VLT / UVES}                                                \\ \hline
075.C-0321(A) & 2005-08-26                         & 200 - 600                \\
082.C-0005(B) & 2008-10-02                         & 10 - 150                 \\ \hline           
\end{tabular}
\label{tab:arc}
\end{table}

\clearpage
\begin{longtable}[htbp]{lrlllllllllll}
\caption{\label{tab:myTeff} Luminosity, temperature, surface gravity, and veiling values of $64\ $PMS stars }\\
\hline\hline
Object Name             & $L_{\rm star}$ & $T_{\rm eff}$ & $\log\ g$ & \multicolumn{9}{c}{Veiling, $r$}    \\
                        & $[L_\odot]$ &  $\mathrm{[K]}$  & $\mathrm{[cm\cdot s^{-2}]}$ & 4400& 5120& 6050& 7020& 7600& 8160& 8500 & 8662 & 8807 \\
\hline
\endfirsthead
\caption{continued.}\\
\hline\hline
Object Name             & $L_{\rm star}$ & $T_{\rm eff}$ & $\log\ g$ & \multicolumn{9}{c}{Veiling, $r$}    \\
                        & $[L_\odot]$ &  $\mathrm{[K]}$  & $\mathrm{[cm\cdot s^{-2}]}$ & 4400& 5120& 6050& 7020& 7600& 8160& 8500 & 8662 & 8807 \\
\hline
\endhead
\hline
\endfoot
\hline
%\multicolumn{2}{@{}l@{}}{\hbox to0pt{\parbox{158mm}{\small }\hss}} 
\endlastfoot
  \multicolumn{3}{l}{Taurus molecular cloud}      &                   &           &      &         &         &         &        &        &        &    \\ \hline
AA Tau                  & 0.06     & 4100 $\pm$ 160  & 4.0   $\pm$ 0.8 & 0.6    &        & 0.9    & 1.0    & 0.4    & 0.5    & 0.6          & 0.9  & 0.5  \\
BP Tau                  & 0.31     & 4000 $\pm$ 180  & 3.5   $\pm$ 0.7 & 1.2    &        & 0.7    & 0.5    & 0.2    & 0.3    & 0.4          & 0.6  & 0.2  \\
CQ Tau                  & 1.90     & 6000 $\pm$ 510  & 2.5   $\pm$ 1.4 & 0.4    & 0.0    & 0.3    & 0.3    & 0.2    & 1.2    & 1.2          & 0.0  & 0.0  \\
DE Tau                  & 0.32     & 3800 $\pm$ 170  & 3.0   $\pm$ 1.1 & 0.4    & 5.0    & 0.3    & 0.0    & 0.0    & 0.0    & 0.1          & 0.4  & 0.0  \\
DF Tau                  & 0.60     & 3700 $\pm$ 850  & 2.5   $\pm$ 0.5 & 0.2    &        & 0.6    & 0.0    & 0.0    & 0.0    & 0.1          & 0.7  & 0.0  \\
DG Tau                  & 0.26     & 4300 $\pm$ 350  & 4.0   $\pm$ 1.4 & 0.7    & 5.0    & 0.8    & 0.1    & 0.9    & 0.5    & 0.4          & 1.2  & 0.4  \\
DK Tau                  & 0.32     & 4200 $\pm$ 170  & 4.0   $\pm$ 1.0 & 0.5    & 1.6    & 0.2    & 0.0    & 0.0    & 0.1    & 0.1          & 0.3  & 0.1  \\
DM Tau                  & 0.10     & 3700 $\pm$ 210  & 3.0   $\pm$ 1.0 & 1.8    & 3.0    & 0.7    & 0.1    & 0.1    & 0.0    & 0.0          & 0.2  & 0.0  \\
DN Tau                  & 0.44     & 3900 $\pm$ 210  & 2.5   $\pm$ 0.9 & 0.2    & 0.7    & 0.1    & 0.0    & 0.0    & 0.0    & 0.2          & 0.1  & 0.0  \\
DR Tau                  & 0.29     & 5200 $\pm$ 1020 & 3.0   $\pm$ 1.3 &        &        & 5.0    & 0.0    & 5.0    & 5.0    & 5.0          & 5.0  & 5.0  \\
FP Tau                  & 0.17     & 3700 $\pm$ 760  & 2.5   $\pm$ 0.9 & 0.3    & 0.6    & 0.2    & 0.0    & 0.0    & 0.0    & 0.0          & 0.3  & 0.0  \\
GM Aur                  & 0.40     & 4300 $\pm$ 240  & 4.0   $\pm$ 0.7 & 0.6    & 1.2    & 0.3    & 0.5    & 0.2    & 0.1    & 0.2          & 0.3  & 0.1  \\
HBC 407                 & 0.22     & 5500 $\pm$ 490  & 5.0   $\pm$ 0.2 & 0.1    & 0.3    & 0.0    & 0.1    & 0.0    & 0.0    & 0.0          & 0.0  & 0.0  \\
HP Tau                  & 0.44     & 4800 $\pm$ 490  & 4.0   $\pm$ 1.2 & 0.3    & 1.2    & 0.5    & 0.0    & 1.4    & 0.1    & 0.0          & 0.6  & 0.3  \\
IQ Tau                  & 0.14     & 4000 $\pm$ 280  & 4.0   $\pm$ 0.9 & 0.6    & 1.8    & 0.4    & 0.0    & 0.1    & 0.1    & 0.2          & 0.5  & 0.2  \\
LkCa 04 (24 Nov)        & 0.29     & 3900 $\pm$ 320  & 3.0   $\pm$ 1.0 & 0.2    & 0.8    & 0.2    & 0.0    & 0.0    & 0.0    & 0.1          & 0.3  & 0.0  \\
LkCa 04 (26 Nov)        & 0.29     & 3900 $\pm$ 280  & 3.0   $\pm$ 0.9 & 0.2    & 0.6    & 0.2    & 0.0    & 0.0    & 0.0    & 0.0          & 0.0  & 0.0  \\
LkCa 15 (27 Feb)        & 0.22     & 4600 $\pm$ 250  & 4.5   $\pm$ 0.8 & 0.4    & 1.0    & 0.1    & 5.0    & 0.0    & 0.3    & 0.1          & 0.4  & 0.3  \\
LkCa 15 (6 Mar)         & 0.56     & 4600 $\pm$ 250  & 4.5   $\pm$ 0.8 & 0.3    & 0.9    & 0.0    & 0.1    & 0.0    & 0.2    & 0.0          & 0.3  & 0.1  \\
RY Tau                  & 4.02     & 5400 $\pm$ 330  & 2.5   $\pm$ 0.8 & 0.2    & 0.1    & 0.2    & 0.6    & 0.0    & 0.1    & 0.1          & 0.3  & 0.0  \\
T Tau                   & 3.59     & 5300 $\pm$ 420  & 4.0   $\pm$ 0.7 & 0.0    & 2.0    & 0.0    & 0.0    & 0.8    & 0.2    & 0.0          & 0.3  & 0.1  \\
UX Tau                  & 1.48     & 5500 $\pm$ 480  & 5.0   $\pm$ 1.2 & 0.2    & 0.3    & 0.0    & 0.0    & 0.0    & 0.2    & 0.0          & 0.0  & 0.0  \\
V1354 Tau               & 0.66     & 5300 $\pm$ 570  & 5.0   $\pm$ 0.3 & 0.1    & 0.3    & 0.0    & 0.1    & 0.0    & 0.0    & 0.0          & 0.0  & 0.0  \\
   &&                                                            &                   &           &      &         &         &         &        &        &        &    \\ \hline
\multicolumn{3}{l}{${\rm\rho}$ Ophiuchi molecular cloud} &        &           &      &         &         &         &        &        &        &    \\ \hline
V2062 Oph               & 1.50     & 4400 $\pm$ 140  & 3.5   $\pm$ 1.0 & 0.4    & 2.5    & 0.1    & 0.3    & 0.4    & 0.2    & 0.2          & 0.4  & 0.2  \\
V2129 Oph               & 0.58     & 4000 $\pm$ 390  & 2.5   $\pm$ 1.0 & 0.3    & 0.7    & 0.2    & 0.2    & 0.0    & 0.0    & 0.4          & 0.3  & 0.0  \\
V2247 Oph               & 0.24     & 3800 $\pm$ 260  & 3.5   $\pm$ 0.6 & 0.4    & 5.0    & 0.4    & 0.0    & 0.0    & 0.0    & 0.0          & 0.4  & 0.0  \\
   &&                                                            &                   &           &      &         &         &         &        &        &        &    \\ \hline
\multicolumn{3}{l}{Lupus star-forming region}       &                   &           &      &         &         &         &        &        &        &    \\ \hline
GQ Lup                  & 1.03     & 4300 $\pm$ 200  & 3.5   $\pm$ 0.7 & 0.6    & 1.6    & 0.3    & 0.6    & 0.2    & 0.3    & 0.4          & 0.6  & 0.2  \\
   &&                                                            &                   &           &      &         &         &         &        &        &        &    \\ \hline
\multicolumn{3}{l}{Chamaeleon I star-forming region}      &        &           &      &         &         &         &        &        &        &    \\ \hline
CHX22                   & 4.35     & 5400 $\pm$ 400  & 4.5   $\pm$ 1.1 & 0.0    & 0.2    & 0.0    & 0.0    & 0.0    & 0.0    & 0.0          & 0.0  & 0.0  \\
CR Cha                  & 2.43     & 5300 $\pm$ 440  & 4.5   $\pm$ 1.2 & 0.0    & 0.4    & 0.0    & 0.0    & 0.0    & 0.0    & 0.0          & 0.1  & 0.1  \\
CV Cha                  & 2.04     & 5000 $\pm$ 290  & 4.0   $\pm$ 1.1 & 0.3    & 3.0    & 0.2    & 0.3    & 0.3    & 0.4    & 0.2          & 0.4  & 0.2  \\
CW Cha                  & 0.03     & 3800 $\pm$ 940  & 4.5   $\pm$ 1.3 & 2.5    & 5.0    & 3.0    & 0.7    & 1.2    & 0.9    & 1.0          & 2.0  & 0.0  \\
   &&                                                            &                   &           &      &         &         &         &        &        &        &    \\ \hline
\multicolumn{3}{l}{Orionis OB 1c association}      &                   &           &      &         &         &         &        &        &        &    \\ \hline
CO Ori                  & 14.88    & 5800 $\pm$ 720  & 4.5   $\pm$ 1.1 & 0.0    & 0.9    & 0.4    & 0.2    & 0.0    & 0.9    & 0.4          & 0.5  & 0.3  \\
COUP 1287               & 0.05     & 3700 $\pm$ 880  & 3.5   $\pm$ 1.4 & 3.0    & 5.0    & 2.5    & 0.0    & 1.0    & 0.4    & 0.0          & 5.0  & 0.0  \\
COUP 1423               & 0.30     & 3600 $\pm$ 120  & 2.0   $\pm$ 1.1 & 0.0    & 0.9    & 0.4    & 0.1    & 0.1    & 0.0    & 0.3          & 0.5  & 0.0  \\
GW Ori                  & 37.29    & 5600 $\pm$ 510  & 4.0   $\pm$ 1.1 & 0.1    & 1.0    & 0.2    & 0.1    & 0.0    & 0.3    & 0.2          & 0.5  & 0.3  \\
HBC 167                 & 7.77     & 5900 $\pm$ 190  & 3.5   $\pm$ 0.9 & 0.0    & 0.3    & 0.0    & 0.1    & 0.2    & 0.1    & 0.0          & 0.7  & 0.0  \\
OV Ori                  & 0.38     & 4300 $\pm$ 350  & 4.0   $\pm$ 0.3 & 0.3    & 0.8    & 0.2    & 0.2    & 0.0    & 0.1    & 0.1          & 0.3  & 0.1  \\
V1044 Ori               & 9.51     & 5800 $\pm$ 660  & 4.5   $\pm$ 1.1 & 0.0    & 0.3    & 0.0    & 0.0    & 0.0    & 0.1    & 0.0          & 0.0  & 0.0  \\
V1458 Ori               & 4.76     & 3800 $\pm$ 340  & 2.5   $\pm$ 0.9 & 0.3    & 0.6    & 0.2    & 0.0    & 0.0    & 0.0    & 0.3          & 0.3  & 0.0  \\
   &&                                                            &                   &           &      &         &         &         &        &        &        &    \\ \hline
\multicolumn{3}{l}{Upper Scorpius association}        &                   &           &      &         &         &         &        &        &        &    \\ \hline
{\tiny 2MASS J16123916-1859284} & 0.13     & 3900 $\pm$ 310  & 4.0   $\pm$ 1.1 & 0.3    & 1.4    & 0.2    & 0.0    & 0.0    & 0.0    & 0.1          & 0.3  & 0.0  \\
V1149 Sco               & 2.81     & 5900 $\pm$ 230  & 4.0   $\pm$ 0.9 & 0.0    & 0.0    & 0.0    & 0.1    & 0.0    & 0.0    & 0.0          & 0.0  & 0.0  \\
   &&                                                            &                   &           &      &         &         &         &        &        &        &    \\ \hline
\multicolumn{3}{l}{AB Doradus moving group}       &                   &           &      &         &         &         &        &        &        &    \\ \hline
HIP 17695               & 0.02     & 3500 $\pm$ 80   & 5.0   $\pm$ 1.3 & 0.1    & 0.7    & 0.1    & 0.0    & 0.0    & 0.1    & 0.1          & 0.4  & 0.0  \\
   &&                                                            &                   &           &      &         &         &         &        &        &        &    \\   \hline 
\multicolumn{3}{l}{${\rm\beta}$ Picoris moving group}      &        &           &      &         &         &         &        &        &        &    \\ \hline
AU Mic                  & 0.05     & 3900 $\pm$ 130  & 4.0   $\pm$ 0.8 & 0.5    &        & 0.2    & 0.0    & 0.0    & 0.0    & 0.0          & 0.0  & 0.0  \\
   &&                                                            &                   &           &      &         &         &         &        &        &        &    \\ \hline
\multicolumn{3}{l}{"Cha-Near" region}                &                   &           &      &         &         &         &        &        &        &    \\ \hline
RX J1147.7-7842         & 0.13     & 3600 $\pm$ 390  & 2.0   $\pm$ 0.8 & 0.1    & 0.6    & 0.1    & 0.0    & 0.0    &        & 0.0          & 0.3  & 0.0  \\
RX J1204.6-7731         & 0.09     & 3500 $\pm$ 190  & 4.5   $\pm$ 0.8 &        &        & 0.3    & 0.0    & 0.2    & 0.1    & 0.0          & 0.2  & 0.0  \\
   &&                                                            &                   &           &      &         &         &         &        &        &        &    \\ \hline
\multicolumn{3}{l}{${\rm\eta}$ Chamaeleontis cluster}     &        &           &      &         &         &         &        &        &        &    \\ \hline
RECX 04                 & 0.14     & 3900 $\pm$ 310  & 4.0   $\pm$ 0.7 & 0.4    & 0.9    & 0.1    & 0.0    & 0.0    & 0.0    & 0.0          & 0.1  & 0.0  \\
RECX 06                 & 0.06     & 3500 $\pm$ 130  & 4.5   $\pm$ 1.2 &        &        & 0.3    & 0.0    & 0.0    & 0.1    & 0.0          & 0.0  & 0.0  \\
RECX 07                 & 0.45     & 4300 $\pm$ 400  & 3.5   $\pm$ 1.1 & 0.2    & 0.6    & 0.0    & 0.0    & 0.1    & 0.1    & 0.3          & 0.3  & 0.2  \\
RECX 09                 & 0.04     & 3000 $\pm$ 380  & 4.0   $\pm$ 1.2 &        &        & 0.3    & 0.0    & 0.2    & 0.3    & 0.2          & 0.5  & 0.0  \\
RECX 10                 & 0.15     & 4100 $\pm$ 270  & 4.0   $\pm$ 0.7 & 0.4    & 0.7    & 0.1    & 0.0    & 0.0    & 0.0    & 0.0          & 0.2  & 0.1  \\
RECX 11                 & 0.34     & 4600 $\pm$ 340  & 4.5   $\pm$ 0.3 & 0.2    & 0.7    & 0.0    & 0.3    & 0.0    & 0.1    & 0.1          & 0.2  & 0.2  \\
RECX 15                 & 0.03     & 3400 $\pm$ 430  & 4.0   $\pm$ 1.4 &        &        & 0.9    & 0.0    & 0.2    & 0.1    & 0.0          & 0.2  & 0.0  \\
   &&                                                            &                   &           &      &         &         &         &        &        &        &    \\ \hline
   \multicolumn{3}{l}{TW Hydrae association}     &                   &           &      &         &         &         &        &        &        &    \\ \hline
TWA 1                   & 0.17     & 4100 $\pm$ 270  & 4.0   $\pm$ 0.6 & 1.0    & 2.0    & 0.5    & 0.0    & 0.2    & 0.1    & 0.2          & 0.4  & 0.3  \\
TWA 5B                  & 0.17     & 3400 $\pm$ 210  & 4.5   $\pm$ 1.3 &        &        & 0.4    & 0.0    & 0.0    & 0.0    & 0.0          & 0.1  & 0.0  \\
TWA 6                   & 0.17     & 4200 $\pm$ 320  & 4.0   $\pm$ 0.6 & 0.1    & 0.7    & 0.0    & 0.0    & 0.0    & 0.0    & 0.0          & 0.0  & 0.1  \\
TWA 7                   & 0.06     & 3500 $\pm$ 180  & 4.5   $\pm$ 0.6 &        &        & 0.3    & 0.0    & 0.0    & 0.0    & 0.0          & 0.0  & 0.0  \\
TWA 8A                  & 0.06     & 3500 $\pm$ 160  & 4.5   $\pm$ 1.1 &        &        & 0.3    & 0.0    & 0.5    & 0.1    & 0.0          & 0.5  & 0.0  \\
TWA 8B                  & 0.01     & 2800 $\pm$ 1130 & 4.0   $\pm$ 1.2 &        &        & 0.4    & 0.0    & 1.4    & 0.3    & 0.3          & 0.6  & 0.0  \\
TWA 9A                  & 0.25     & 4000 $\pm$ 290  & 3.0   $\pm$ 0.6 & 0.1    & 0.6    & 0.0    & 0.1    & 0.0    & 0.0    & 0.0          & 0.0  & 0.0  \\
TWA 9B                  & 0.04     & 3400 $\pm$ 470  & 4.5   $\pm$ 0.6 &        &        & 0.2    & 0.0    & 0.7    & 0.1    & 0.0          & 0.1  & 0.0  \\
TWA 14                  & 0.14     & 3900 $\pm$ 360  & 4.0   $\pm$ 1.0 & 0.3    & 0.8    & 0.3    & 0.0    & 0.0    & 0.0    & 0.0          & 0.1  & 0.0  \\
TWA 22                  & 0.01     & 2800 $\pm$ 320  & 4.5   $\pm$ 0.5 &        &        & 0.2    & 0.0    & 0.3    &        & 0.2          & 0.7  & 0.0  \\
TWA 25                  & 0.17     & 4000 $\pm$ 320  & 3.5   $\pm$ 0.8 & 0.2    & 0.8    & 0.0    &        & 0.0    & 0.0    & 0.0          & 0.0  & 0.0 \\ \hline
\end{longtable}

\begin{longtable}[tbph]{lcccrrc}
\caption{\label{tab:objects2} Projected rotational velocity ($v\sin\ i$) derived by the previous study, period measured from TESS light curves, stellar age, mass, convective turnover time ($\tau_{\rm c}$), and Rossby number ($N_{\rm R}$) of the $64\ $PMS stars.}\\
\hline\hline
Object Name & $v\sin\ i$ & Period & Age  & Mass & $\tau_{\rm c}$ & $\log\ N_{\rm R}$ \\
                        & $\mathrm{[km\cdot s^{-1}]}$ & $\mathrm{[d]}$ & $\mathrm{[Myr]}$  & $\mathrm{[M_\odot]}$ & $\mathrm{[d]}$ &        \\
\hline
\endfirsthead
\caption{continued.}\\
\hline\hline
Object Name & $v\sin\ i$ & Period & Age  & Mass & $\tau_{\rm c}$ & $\log\ N_{\rm R}$ \\
                        & $\mathrm{[km\cdot s^{-1}]}$ & $\mathrm{[d]}$ & $\mathrm{[Myr]}$  & $\mathrm{[M_\odot]}$ & $\mathrm{[d]}$ &        \\
\hline
\endhead
\hline
\endfoot
\hline
\multicolumn{2}{@{}l@{}}{\hbox to0pt{\parbox{128mm}{\small
$v \sin i$; the projected rotational velocity from the Catalog of Stellar Rotational Velocities \citep{g05}. (*) The rotational periods are estimated with their $v\sin\ i$ because TESS light curve were not obtained.
}\hss}} 
\endlastfoot
\multicolumn{3}{l}{Taurus molecular cloud}  &      &      &      &        \\ \hline
AA Tau                  & 12.7  & 7.03   & 7.9 & 0.53 & 75   & -1.03  \\
BP Tau                  & 11.4  & 7.18   & 6.5 & 0.53 & 252  & -1.55  \\
CQ Tau                  & 107.5 & 12.16  & 7.3 & 1.19 & 22   & -0.26  \\
DE Tau                  & 10.0  & 5.69   & 6.2 & 0.39 & 232  & -1.61  \\
DF Tau                  & 19.1  & 10.63  & 5.7 & 0.34 & 225  & -1.33  \\
DG Tau                  & 21.7  & 17.99  & 7.2 & 0.82 & 102  & -0.75  \\
DK Tau                  & 11.4  & 6.56   & 6.9 & 0.78 & 161  & -1.39  \\
DM Tau                  & 10.0  & 7.32   & 6.9 & 0.35 & 240  & -1.52  \\
DN Tau                  & 9.4   & 6.49   & 6.1 & 0.44 & 234  & -1.56  \\
DR Tau                  & 10.0  & 16.47  & 7.7 & 0.81 & 31   & -0.28  \\
FP Tau                  & 23.5  & 2.18   & 6.5 & 0.34 & 232  & -2.03  \\
GM Aur                  & 13.0  & 5.43   & 6.9 & 0.87 & 143  & -1.42  \\
HBC 407                 & 10.0  & 3.45   & 7.8 & 0.76 & 35   & -1.00  \\
HP Tau                  & 66.0  & 0.44*   & 7.3 & 0.94 & 57   & -1.88  \\
IQ Tau                  & 12.0  & 6.63   & 7.2 & 0.58 & 172  & -1.41  \\
LkCa 04 (24 Nov)        & 26.1  & 3.37   & 6.5 & 0.48 & 244  & -1.86  \\
LkCa 04 (26 Nov)        & 26.1  & 3.37   & 6.5 & 0.48 & 244  & -1.86  \\
LkCa 15 (27 Feb)        & 12.5  & 5.13   & 7.6 & 0.76 & 51   & -0.99  \\
LkCa 15 (6 Mar)         & 12.5  & 5.13   & 7.0 & 1.04 & 86   & -1.22  \\
RY Tau                  & 50.8  & 5.24   & 6.6 & 1.83 & 63   & -1.08  \\
T Tau                   & 21.0  & 14.96  & 6.6 & 1.79 & 69   & -0.66  \\
UX Tau                  & 20.7  & 6.42   & 7.1 & 1.22 & 63   & -0.99  \\
V1354 Tau               &       & 1.46   & 7.5 & 0.94 & 32   & -1.34  \\
                            &       &        &      &      &      &        \\ \hline
\multicolumn{3}{l}{${\rm\rho}$ Ophiuchi molecular cloud}  &      &      &      &  \\ \hline
V2062 Oph               & 8.5   & 5.79*   & 6.1 & 0.90 & 266  & -1.33  \\
V2129 Oph               & 15.8  & 5.11*   & 6.1 & 0.52 & 240  & -1.67  \\
V2247 Oph               & 43.1  & 1.11*   & 6.4 & 0.40 & 236  & -2.25  \\
                             &       &        &      &      &      &        \\ \hline
\multicolumn{3}{l}{Lupus star-forming region}  &      &      &      &        \\  \hline
GQ Lup                  &       & 10.77  & 6.2 & 0.78 & 261  & -1.38  \\
                             &       &        &      &      &      &        \\ \hline
\multicolumn{3}{l}{Chamaeleon I star-forming region}  &      &      &      &        \\ \hline
CHX22                   & 11.3  & 7.37*   & 6.6 & 1.87 & 62   & -0.77  \\
CR Cha                  & 36.7  & 2.21*   & 6.8 & 1.56 & 57   & -1.35  \\
CV Cha                  & 30.3  & 3.00*   & 6.6 & 1.53 & 93   & -1.47  \\
CW Cha                  & 29.6  & 1.06*   & 7.8 & 0.43 & 117  & -2.20  \\
                             &       &        &      &      &      &        \\ \hline
\multicolumn{3}{l}{Orionis OB 1c association}  &      &      &      &        \\ \hline
CO Ori                  & 56.6  & 2.32*   & 6.3 & 2.55 & 34   & -0.99  \\
COUP 1287               & 32.7  & 0.78*   & 7.4 & 0.36 & 226  & -2.44  \\
COUP 1423               & 40.7  & 1.62*   & 6.0 & 0.29 & 225  & -2.11  \\
GW Ori                  & 41.7  & 5.67*   & 5.8 & 3.56 & 31   & -0.59  \\
HBC 167                 & 18.0  & 7.52*   & 6.7 & 1.91 & 74   & -0.99  \\
OV Ori                  & 20.1  & 2.31*   & 6.9 & 0.88 & 133  & -1.67  \\
V1044 Ori               & 22.3  & 4.66*   & 6.5 & 2.23 & 31   & -0.65  \\
V1458 Ori               & 20.3  & 11.31*  & 4.3 & 0.36 & 217  & -1.24  \\
                            &       &        &      &      &      &        \\ \hline
\multicolumn{3}{l}{Upper Scorpius association}  &      &      &      &        \\ \hline
2MASS J16123916-1859284 & 12.3  & 3.00*   & 7.0 & 0.49 & 248  & -1.87  \\
V1149 Sco               &       &        & 7.1 & 1.40 & 36   &        \\
                             &       &        &      &      &      &        \\  \hline
\multicolumn{3}{l}{AB Doradus moving group}  &      &      &      &        \\ \hline
HIP 17695               &       & 3.86   & 7.7 & 0.26 & 252  & -1.82  \\
                             &       &        &      &      &      &        \\ \hline
\multicolumn{3}{l}{${\rm\beta}$ Picoris moving group}  &&&& \\ \hline
AU Mic                  & 8.0   & 4.80   & 7.7 & 0.51 & 99   & -1.31  \\
                             &       &        &      &      &      &        \\ \hline
\multicolumn{3}{l}{"Cha-Near" region}  &      &      &      &        \\ \hline
RX J1147.7-7842         &       &        & 6.6 & 0.30 & 229  &        \\
RX J1204.6-7731         & 9.5   & 4.79   & 6.6 & 0.25 & 225  & -1.67  \\
                              &       &        &      &      &      &        \\ \hline
\multicolumn{3}{l}{${\rm\eta}$ Chamaeleontis cluster} &      &      &      &        \\ \hline
RECX 04                 &       & 6.97   & 7.0 & 0.50 & 245  & -1.55  \\
RECX 06                 &       & 1.84   & 6.9 & 0.25 & 227  & -2.09  \\
RECX 07                 & 30.0  & 1.99*   & 6.8 & 0.88 & 155  & -1.88  \\
RECX 09                 &       & 1.94   & 6.0 & 0.08 & 186  & -1.98  \\
RECX 10                 & 11.0  & 8.74   & 7.3 & 0.68 & 108  & -1.09  \\
RECX 11                 &       & 4.98   & 7.4 & 0.87 & 62   & -1.09  \\
RECX 15                 & 30.0  & 0.98*   & 7.0 & 0.21 & 222  & -2.39  \\
                              &       &        &      &      &      &        \\ \hline
\multicolumn{3}{l}{TW Hydrae association} &      &      &      &        \\ \hline
TWA 1                   & 15.0  & 7.08   & 7.2 & 0.67 & 138  & -1.29  \\
TWA 5B                  & 59.0  & 0.73   & 6.2 & 0.25 & 224  & -2.49  \\
TWA 6                   &       & 0.54   & 7.3 & 0.73 & 97   & -2.25  \\
TWA 7                   & 63.2  & 4.92   & 6.8 & 0.25 & 227  & -1.66  \\
TWA 8A                  & 5.0   & 4.63   & 6.8 & 0.25 & 226  & -1.69  \\
TWA 8B                  & 5.0   & 4.73   & 7.1 & 0.07 & 180  & -1.58  \\
TWA 9A                  &       & 5.03   & 6.7 & 0.55 & 266  & -1.72  \\
TWA 9B                  &       & 4.96   & 7.0 & 0.21 & 222  & -1.65  \\
TWA 14                  &       & 0.63   & 7.0 & 0.49 & 252  & -2.60  \\
TWA 22                  &       & 0.73   & 7.0 & 0.07 & 179  & -2.39  \\
TWA 25                  & 12.9  & 5.05   & 7.0 & 0.59 & 192  & -1.58  \\ \hline
\end{longtable}

\begin{figure}[htbp]
\begin{center}
    %\vspace{-2.5cm}
    \resizebox{\hsize}{!}{\includegraphics[clip, width=17cm]{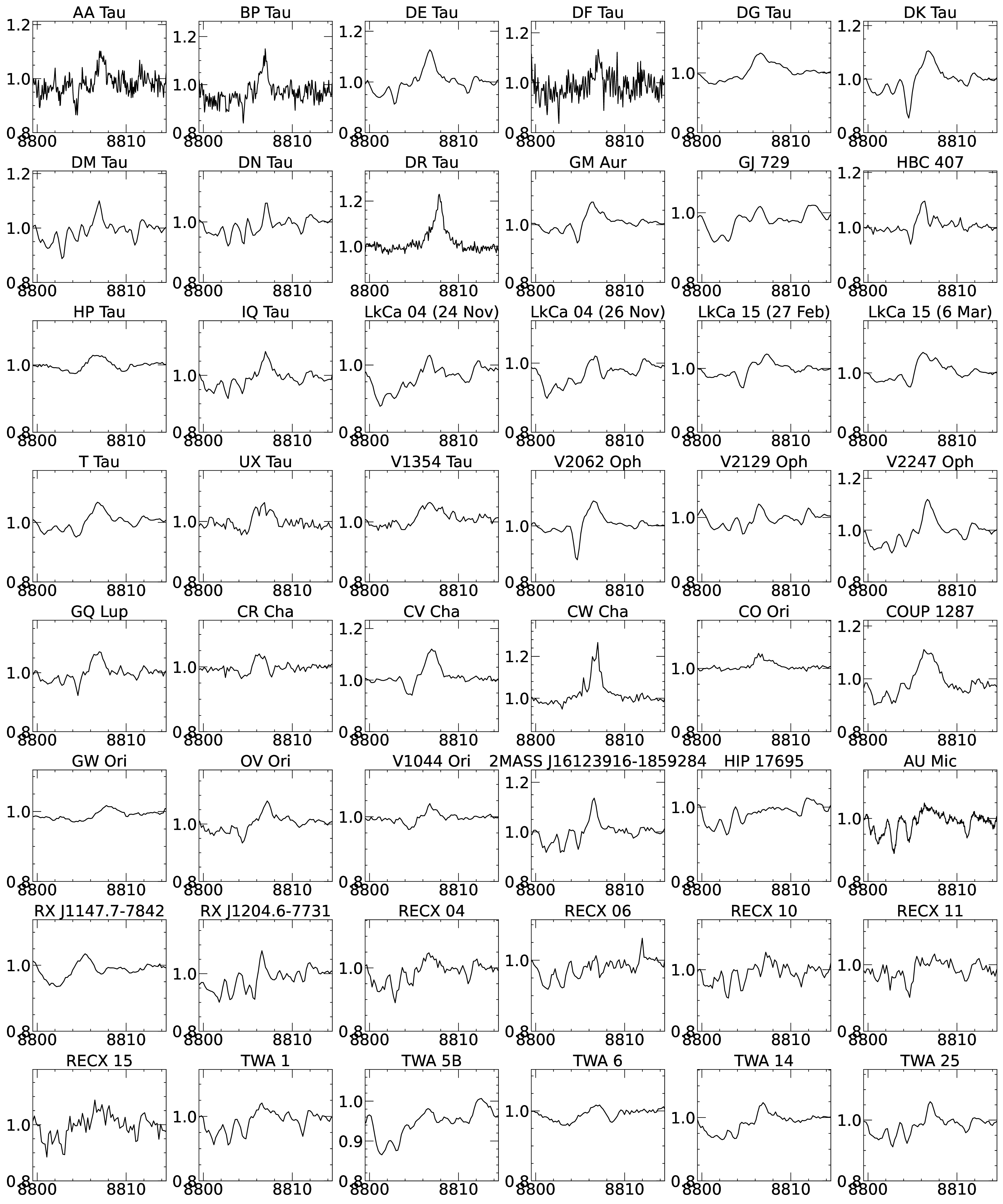}}
    %\vspace{-3.5cm}
    \caption{Mg I ($8807 \  \mathrm{\AA}$) emission lines of the $48\ $PMS stars. Continuum is normalized to unity. Photospheric absorption lines are already subtracted.}
    \label{fig:result_emi2P}
\end{center}
\end{figure}

\begin{figure}[hbpt]
\begin{center}
    %\vspace{-2.5cm}
    \resizebox{\hsize}{!}{\includegraphics[clip, width=17cm]{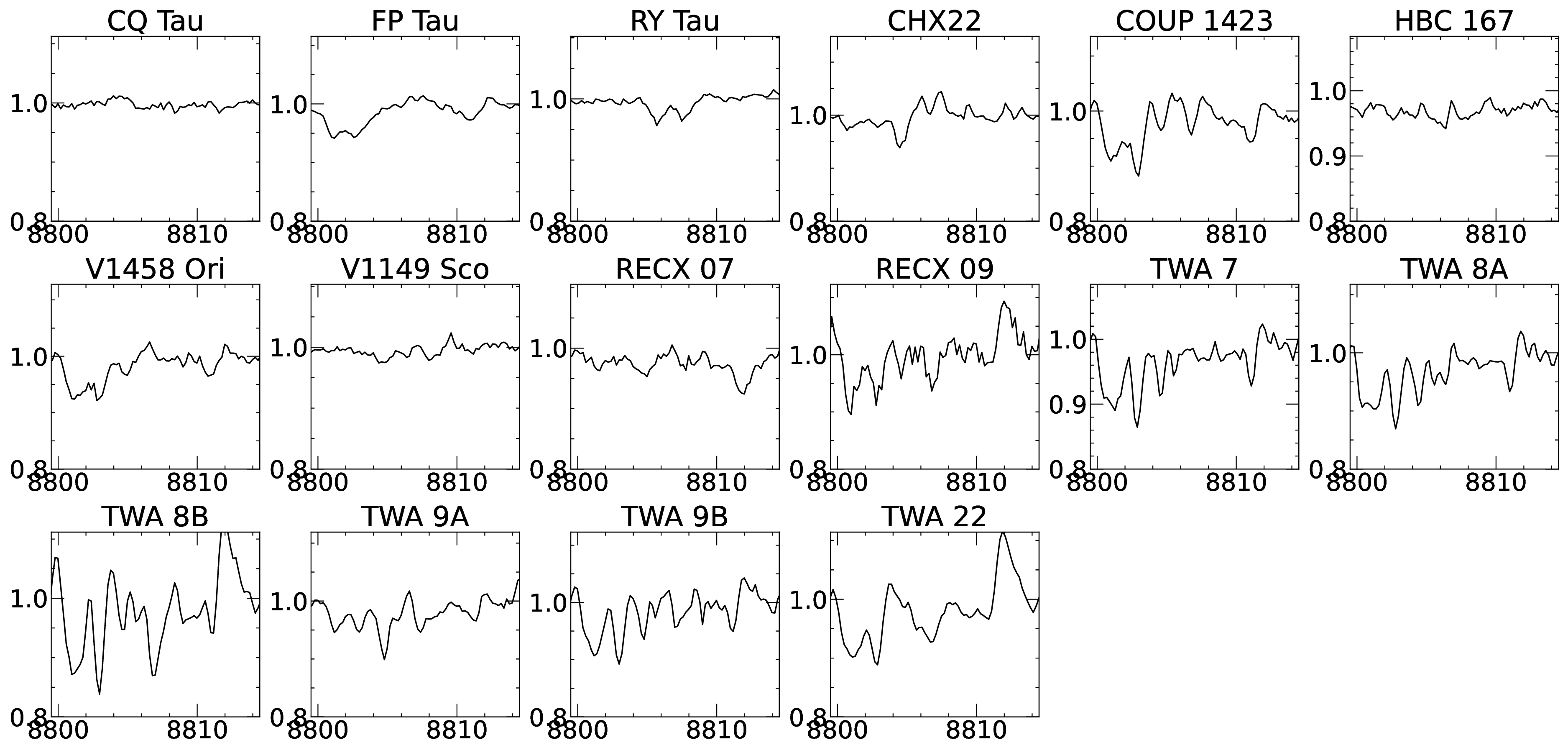}}
    %\vspace{-3.5cm}
    \caption{Spectra near Mg I ($8807\ \mathrm{\AA}$) emission lines of the $16\ $PMS stars without significant detection. Continuum is normalized to unity. Photospheric absorption lines are already subtracted.}
    \label{fig:result_emi2Pn}
\end{center}
\end{figure}

\clearpage
\begin{longtable}[htbp]{lcccccc}
\caption{\label{tab:result_of_obs2} EQWs, FWHMs, and $R^{\prime}$ of the chromospheric emission lines of Ca II IRT ($8542\ \mathrm{\AA}$), and Mg I ($8807\ \mathrm{\AA}$). The EQW values in parentheses indicate non-significant. }\\
\hline\hline
%Object Name & \multicolumn{2}{c}{EQW$\ \mathrm{[\AA]}$}    & \multicolumn{2}{c}{FWHM$\ \mathrm{[km\cdot s^{-1}]}$}   & \multicolumn{2}{c}{$\log\ R^{\prime}$} \\ 
%                       & Ca II & Mg I & Ca II & Mg I & Ca II & Mg I \\
%\hline
\endfirsthead
\caption{continued.}\\
\hline\hline
Object Name & \multicolumn{2}{c}{EQW$\ \mathrm{[\AA]}$}    & \multicolumn{2}{c}{FWHM$\ \mathrm{[km\cdot s^{-1}]}$}   & \multicolumn{2}{c}{$\log\ R^{\prime}$} \\ 
                       & Ca II & Mg I & Ca II & Mg I & Ca II & Mg I \\
\hline
\endhead
\hline
\endfoot
\hline
%\multicolumn{2}{@{}l@{}}{\hbox to0pt{\parbox{158mm}{\small
%memo memo memo
%}\hss}} 
\endlastfoot
Object Name & \multicolumn{2}{c}{EQW$\ \mathrm{[\AA]}$}    & \multicolumn{2}{c}{FWHM$\ \mathrm{[km\cdot s^{-1}]}$}   & \multicolumn{2}{c}{$\log\ R^{\prime}$} \\ 
                       & Ca II & Mg I & Ca II & Mg I & Ca II & Mg I \\ \hline
\multicolumn{3}{l}{Taurus molecular cloud} &&&& \\ \hline
AA Tau                  & -1.32 $\pm$0.10 & -0.11$\pm$0.04 & 26    & 30    & -3.61$\pm$0.01 & -4.68$\pm$0.15\\
BP Tau                  & -1.67 $\pm$0.26 & -0.15$\pm$0.05 & 31    & 35    & -3.60$\pm$0.01 & -4.65$\pm$0.14\\
CQ Tau                  &       -     &      - &    -   &    -   &      -     &      -    \\
DE Tau                  & -9.61 $\pm$0.14 & -0.24$\pm$0.07 & 129   & 48    & -2.97$\pm$0.00 & -4.56$\pm$0.12\\
DF Tau                  & -2.14 $\pm$0.24 & -0.10$\pm$0.07 & 61    & 49    & -3.62$\pm$0.01 & -4.97$\pm$0.31\\
DG Tau                  & -16.71$\pm$0.13 & -0.21$\pm$0.02 & 180   & 104   & -2.55$\pm$0.00 & -4.46$\pm$0.05\\
DK Tau                  & -2.33 $\pm$0.27 & -0.33$\pm$0.05 & 48    & 61    & -3.57$\pm$0.01 & -4.42$\pm$0.06\\
DM Tau                  & -1.35 $\pm$0.16 & -0.13$\pm$0.03 & 29    & 30    & -3.79$\pm$0.01 & -4.81$\pm$0.10\\
DN Tau                  & -1.38 $\pm$0.04 & -0.06$\pm$0.02 & 32    & 18    & -3.81$\pm$0.01 & -5.15$\pm$0.13\\
DR Tau                  & -42.71$\pm$0.25 & -0.44$\pm$0.05 & 148   & 59    & -1.73$\pm$0.00 & -3.72$\pm$0.05\\
FP Tau                  & -0.89 $\pm$0.05 & (-0.04$\pm$0.04) & 67    &    -   & -4.00$\pm$0.02 & -5.40$\pm$0.53\\
GM Aur                  & -2.01 $\pm$0.07 & -0.11$\pm$0.03 & 46    & 59    & -3.55$\pm$0.01 & -4.80$\pm$0.13\\
HBC 407                 & -0.47 $\pm$0.03 & -0.10$\pm$0.03 & 27    & 30    & -4.33$\pm$0.02 & -5.02$\pm$0.11\\
HP Tau                  & -0.26 $\pm$0.04 & -0.15$\pm$0.03 & 154   & 88    & -4.39$\pm$0.04 & -4.64$\pm$0.08\\
IQ Tau                  & -1.58 $\pm$0.06 & -0.13$\pm$0.05 & 37    & 42    & -3.72$\pm$0.01 & -4.80$\pm$0.18\\
LkCa 04 (24 Nov)        & -1.36 $\pm$0.06 & -0.09$\pm$0.06 & 56    & 24    & -3.81$\pm$0.02 & -5.01$\pm$0.32\\
LkCa 04 (26 Nov)        & -0.89 $\pm$0.05 & -0.08$\pm$0.04 & 52    & 33    & -4.00$\pm$0.02 & -5.04$\pm$0.23\\
LkCa 15 (27 Feb)        & -1.41 $\pm$0.04 & -0.11$\pm$0.03 & 54    & 51    & -3.37$\pm$0.01 & -4.48$\pm$0.14\\
LkCa 15 (6 Mar)         & -1.24 $\pm$0.08 & -0.19$\pm$0.05 & 42    & 100   & -3.83$\pm$0.02 & -4.64$\pm$0.11\\
RY Tau                  & -3.63 $\pm$0.07 & (-0.03$\pm$0.04) & 164   &    -   & -3.36$\pm$0.00 & -5.48$\pm$0.56\\
T Tau                   & -10.87$\pm$0.14 & -0.13$\pm$0.05 & 110   & 54    & -2.85$\pm$0.00 & -4.77$\pm$0.18\\
UX Tau                  & -0.78 $\pm$0.05 & -0.15$\pm$0.05 & 49    & 73    & -4.09$\pm$0.03 & -4.81$\pm$0.14\\
V1354 Tau               & -0.65 $\pm$0.05 & -0.19$\pm$0.04 & 53    & 90    & -4.18$\pm$0.03 & -4.71$\pm$0.09\\
 &&&&&& \\ \hline
\multicolumn{3}{l}{${\rm\rho}$ Ophiuchi molecular cloud} &&&& \\ \hline
V2062 Oph               & -13.26$\pm$0.26 & -0.11$\pm$0.02 & 151   & 49    & -2.72$\pm$0.00 & -4.79$\pm$0.07\\
V2129 Oph               & -1.26 $\pm$0.05 & -0.05$\pm$0.02 & 34    & 32    & -3.82$\pm$0.01 & -5.18$\pm$0.19\\
V2247 Oph               & -2.84 $\pm$0.08 & -0.17$\pm$0.04 & 41    & 42    & -3.50$\pm$0.01 & -4.71$\pm$0.09\\
 &&&&&& \\ \hline
\multicolumn{3}{l}{Lupus star-forming region} &&&& \\ \hline
GQ Lup                  & -3.18 $\pm$0.11 & -0.12$\pm$0.04 & 52    & 49    & -3.31$\pm$0.01 & -4.75$\pm$0.15\\
 &&&&&& \\ \hline
\multicolumn{3}{l}{Chamaeleon I star-forming region} &&&& \\ \hline
CHX22                   & -0.63 $\pm$0.04 & (-0.08$\pm$0.08) & 29    & 23    & -4.21$\pm$0.06 & -5.12$\pm$0.45\\
CR Cha                  & -0.97 $\pm$0.05 & -0.12$\pm$0.04 & 72    & 51    & -4.02$\pm$0.02 & -4.92$\pm$0.14\\
CV Cha                  & -16.30$\pm$1.36 & -0.18$\pm$0.03 & 240   & 51    & -2.65$\pm$0.00 & -4.62$\pm$0.08\\
CW Cha                  & -7.57 $\pm$0.16 & -0.53$\pm$0.10 & 51    & 47    & -2.78$\pm$0.01 & -3.94$\pm$0.08\\
 &&&&&& \\ \hline
\multicolumn{3}{l}{Orionis OB 1c association} &&&& \\ \hline
CO Ori                  & -1.98 $\pm$0.20 & -0.08$\pm$0.03 & 144   & 83    & -3.60$\pm$0.01 & -5.02$\pm$0.20\\
COUP 1287               & -4.98 $\pm$0.23 & -0.45$\pm$0.10 & 121   & 79    & -3.09$\pm$0.01 & -4.13$\pm$0.09\\
COUP 1423               & -0.96 $\pm$0.05 &      - & 36    & 40    & -3.94$\pm$0.02 &      -    \\
GW Ori                  & -4.59 $\pm$0.10 & -0.05$\pm$0.03 & 176   & 67    & -3.31$\pm$0.00 & -5.29$\pm$0.27\\
HBC 167                 & -3.11 $\pm$0.07 &      - & 86    &   -    & -3.49$\pm$0.00 &      -    \\
OV Ori                  & -1.24 $\pm$0.05 & -0.12$\pm$0.04 & 44    & 45    & -3.82$\pm$0.01 & -4.82$\pm$0.14\\
V1044 Ori               & -2.74 $\pm$0.07 & -0.07$\pm$0.04 & 95    & 44    & -3.58$\pm$0.01 & -5.21$\pm$0.26\\
V1458 Ori               & -0.60 $\pm$0.04 & (-0.02$\pm$0.07) & 39    &   -    & -4.17$\pm$0.05 & -5.57$\pm$1.22\\
 &&&&&& \\ \hline
 \multicolumn{3}{l}{Upper Scorpius association} &&&& \\ \hline
2MASS J16123916-1859284 & -1.22 $\pm$0.13 & -0.13$\pm$0.04 & 31    & 27    & -3.86$\pm$0.02 & -4.83$\pm$0.15\\
V1149 Sco               & -0.85 $\pm$0.05 & (-0.01$\pm$0.04) & 57    &   -    & -4.10$\pm$0.02 & -6.01$\pm$1.50\\
 &&&&&& \\ \hline
\multicolumn{3}{l}{AB Doradus moving group} &&&& \\ \hline
HIP 17695               & -0.40 $\pm$0.03 & -0.09$\pm$0.05 & 31    & 101   & -4.35$\pm$0.06 & -4.99$\pm$0.25\\
 &&&&&& \\ \hline
\multicolumn{3}{l}{${\rm\beta}$ Picoris moving group}  &&&& \\ \hline
AU Mic                  & -0.69 $\pm$0.05 & -0.24$\pm$0.06 & 21    & 74    & -4.11$\pm$0.04 & -4.58$\pm$0.11\\
 &&&&&& \\ \hline
 \multicolumn{3}{l}{"Cha-Near" region} &&&& \\ \hline
RX J1147.7-7842         & -0.84 $\pm$0.07 & -0.11$\pm$0.06 & 144   & 36    & -4.03$\pm$0.03 & -4.91$\pm$0.23\\
RX J1204.6-7731         & -0.65 $\pm$0.07 & -0.07$\pm$0.02 & 27    & 21    & -4.11$\pm$0.02 & -5.11$\pm$0.16\\
 &&&&&& \\ \hline
\multicolumn{3}{l}{${\rm\eta}$ Chamaeleontis cluster} &&&& \\ \hline
RECX 04                 & -0.94 $\pm$0.06 & -0.28$\pm$0.03 & 28    & 54    & -3.98$\pm$0.01 & -4.50$\pm$0.04\\
RECX 06                 & -0.60 $\pm$0.05 & -0.06$\pm$0.03 & 35    & 21    & -4.17$\pm$0.02 & -5.18$\pm$0.22\\
RECX 07                 & -0.70 $\pm$0.03 & (-0.06$\pm$0.07) & 49    &   -    & -4.08$\pm$0.04 & -5.14$\pm$0.47\\
RECX 09                 & -0.20 $\pm$0.10 &      - & 37    &    -   & -4.66$\pm$0.15 &      -    \\
RECX 10                 & -0.70 $\pm$0.10 & -0.10$\pm$0.03 & 27    & 38    & -4.10$\pm$0.02 & -4.96$\pm$0.13\\
RECX 11                 & -0.87 $\pm$0.07 & -0.29$\pm$0.07 & 31    & 81    & -3.98$\pm$0.04 & -4.45$\pm$0.11\\
RECX 15                 & -9.09 $\pm$0.66 & -0.17$\pm$0.10 & 230   & 71    & -2.97$\pm$0.00 & -4.71$\pm$0.25\\
 &&&&&& \\ \hline
\multicolumn{3}{l}{TW Hydrae association}  &&&& \\ \hline
TWA 1                  & -1.64 $\pm$0.11 & -0.34$\pm$0.04 & 34    & 56    & -3.69$\pm$0.01 & -4.38$\pm$0.06\\
TWA 5B                  & -0.82 $\pm$0.08 & -0.04$\pm$0.03 & 93    & 34    & -4.05$\pm$0.02 & -5.31$\pm$0.31\\
TWA 6                  & -0.90 $\pm$0.05 & -0.14$\pm$0.02 & 104   & 84    & -4.00$\pm$0.01 & -4.82$\pm$0.07\\
TWA 7                  & -0.69 $\pm$0.07 & (0.00 $\pm$0.02) & 24    &    -   & -4.12$\pm$0.02 & -6.57$\pm$4.31\\
TWA 8A                  & -0.96 $\pm$0.06 & (-0.01$\pm$0.04) & 22    &    -   & -3.90$\pm$0.02 & -5.72$\pm$1.17\\
TWA 8B                  & (-0.07 $\pm$0.13) & (-0.02$\pm$0.10) &   -    &  -    & -5.03$\pm$0.59 & -5.65$\pm$2.50\\
TWA 9A                  & -0.81 $\pm$0.05 & (-0.01$\pm$0.04) & 28    &   -    & -4.03$\pm$0.02 & -6.00$\pm$2.13\\
TWA 9B                  & -0.51 $\pm$0.08 &      - & 25    &   -    & -4.16$\pm$0.04 &      -    \\
TWA 14                  & -1.02 $\pm$0.06 & -0.12$\pm$0.04 & 72    & 42    & -3.94$\pm$0.02 & -4.87$\pm$0.13\\
TWA 22                  &       -     &      - &   -    &    -   &      -     &      -    \\
TWA 25                  & -0.94 $\pm$0.05 & -0.09$\pm$0.02 & 30    & 27    & -3.97$\pm$0.01 & -5.02$\pm$0.10\\ \hline
\end{longtable}
\twocolumn
\end{appendix}
\end{document}